\documentclass[aps,amsmath,twocolumn,amssymb,floatfix,showpacs,superscriptaddress,nofootinbib,longbibliography]{revtex4-1}
\usepackage{mathtools}
\usepackage{braket}
\usepackage[dvipsnames]{xcolor}
\usepackage{float}
\usepackage{subfigure}
\usepackage{graphics}
\usepackage{bm}
\usepackage{amsmath,amssymb,bm}
\usepackage{amsfonts}

\usepackage{MnSymbol}
\usepackage{braket}
\usepackage{float}
\usepackage{tikz}
\usepackage{blkarray}
\usepackage{pgffor}
\usepackage{float}
\usepackage{silence}
\usepackage[colorlinks=true,linktoc=page,linkcolor=BrickRed,citecolor=magenta,urlcolor=purple]{hyperref}

\mathchardef\mhyphen="2D 

\newcommand{\etal}{{\it et al.}}

\newcommand\bea{\begin{eqnarray}}
\newcommand\eea{\end{eqnarray}}
\newcommand\beq{\begin{equation}}  
\newcommand\eeq{\end{equation}}

\usepackage[normalem]{ulem}
\definecolor{lime}{HTML}{A6CE39}
\usepackage{sidecap,tikz}
\DeclareRobustCommand{\orcidicon}{\hspace{-1.0mm}
	\begin{tikzpicture}
		\draw[lime, fill=lime] (0.0,0.0) 
		circle [radius=0.15] 
		node[white] {{\fontfamily{qag}\selectfont \tiny \,ID}};
		\draw[white, fill=white] (-0.0525,0.095) 
		circle [radius=0.007];
	\end{tikzpicture}
	\hspace{-3.0mm}
}
\foreach \x in {A, ..., Z}{\expandafter\xdef\csname orcid\x\endcsname{\noexpand\href{https://orcid.org/\csname orcidauthor\x\endcsname}
		{\noexpand\orcidicon}}
}

\AtBeginDocument{%
	\newwrite\bibnotes
	\def\bibnotesext{Notes.bib}
	\immediate\openout\bibnotes=\jobname\bibnotesext
	\immediate\write\bibnotes{@CONTROL{REVTEX41Control}}
	\immediate\write\bibnotes{@CONTROL{%
			apsrev41Control,author="08",editor="1",pages="1",title="1",year="1"}}
	\if@filesw
	\immediate\write\@auxout{\string\citation{apsrev41Control}}%
	\fi
}%
\begin{document}
\newcommand{\Z}{\mathbb{Z}}

\title{Holstein polaron in a pseudospin-$1$ quantum spin Hall system: first and second order topological phase transitions}

\author{Kuntal Bhattacharyya$^\S$\orcidA{}}
\email[]{kuntalphy@iitg.ac.in}

\author{Srijata Lahiri$^\S$\orcidB{}}
\email[]{srijata.lahiri@iitg.ac.in} 

\author{Mijanur Islam$^\S$\orcidC{}}
\email[]{mislam@iitg.ac.in} 

\author{Saurabh Basu}
\email[]{saurabh@iitg.ac.in}
\affiliation{Department of Physics, Indian Institute of Technology Guwahati, Guwahati-781039, Assam, India}

\begin{abstract}
\noindent We theoretically propose the occurrence of a quantum spin Hall (QSH) and a second order topological phase transition (TPT) driven by electron-phonon (e-p) coupling in a pseudospin-$1$ fermionic system on an $\alpha$-$T_3$ lattice. Our model is formulated in the spirit of the Kane-Mele model modified by the Holstein Hamiltonian (accounting for the e-p interaction). To investigate the effects of the e-p coupling ($\lambda$) on the spectral properties associated with $\mathbb{Z}_2$ invariant, we employ the Lang-Firsov approach which describes polarons reasonably well in the anti-adiabatic (high frequency) limit and transforms the original Hamiltonian into an effective electronic Hamiltonian followed by a zero phonon averaging at $T=0$. It is shown that the system possesses topologically nontrivial phases up to a critical e-p coupling, $\lambda_c$ and are characterized by the helical QSH edge states along with a non-zero $\mathbb{Z}_2$ invariant for a certain range of $\alpha$. The topological phase vanishes beyond $\lambda_c$ and is accompanied by a bulk gap closing transition at $\lambda_c$, manifesting a topological-trivial TPT. We observe a more intriguing phenomenon for higher values of $\alpha$, in a certain regime, where the system exhibits a trivial-topological-trivial TPT supported by two distinct gap closing transitions at $\lambda_{c_1}$ and $\lambda_{c_2}$, while a slim region at slightly lower values hosts a semi-metallic signature below $\lambda_{c_1}$. Subsequently, to explore more intricate features, we introduce a time reversal symmetry breaking magnetic field to trigger the formation of a second order topological phase. The magnetic field, by construction causes a boundary dependent gapping out of the edge states, consequently giving rise to robust corner modes in a tailored open boundary conditions.  We justify the formation of the higher order phase by employing an appropriate invariant, namely the projected spin Chern number. Finally, we show that the e-p coupling significantly influences the corner modes (and also the real space energy bandstructure), corroborating a higher order TPT as we tune $\lambda$ beyond a critical value for a given value of $\alpha$.

\end{abstract}
\maketitle

\section{Introduction}\label{textintro}  
 
Thanks to the pioneering work by Thouless~\etal~\cite{Thouless1982} giving an account of a topological invariant to the understanding of the integer quantum Hall effect (QHE), interest in engineering topological phases and relative phase transitions~\cite{Hasan2010,Qi2011} remains unabated in two-dimensional (2D)~\cite{Kane2005,Bernevig2006} and three-dimensional (3D) ~\cite{Moore2007,Fu2007, Konig2007,Hsieh2008} materials. Moreover, the discovery of symmetry-protected topological phases has stimulated enormous attention in such systems, where a continuous phase transition is possible between states with same symmetry but different topology~\cite{Hasan2010,Qi2011,Kosterlitz1973,Senthil2015}. The nontrivial features of the first order  topological insulators (FOTIs) are characterized by a $d$-dimensional insulating bulk and a $(d-1)$-dimensional gapless surface or edge which are topologically robust against disorder. These FOTIs are classified into several categories, each of which is designated by a distinct topological index. One of them is Chern insulator, where the topological invariant known as the Chern number ($C$), hosts chiral edge modes. It was first predicted by Haldane~\cite{Haldane1988} that the Chern insulator can be realized in an exotic 2D lattice model that describes quantized Hall plateaus (with a nonzero $C$) even in the absence of an external magnetic flux~\cite{Nagaosa2010,Chang2013,Chang2015,Deng2020}, and the phenomenon is known as the anomalous QHE. Unlike the Chern insulators, which validate the breaking of time reversal symmetry (TRS) to be the origin of topological phenomena, there exists another class of FOTIs, called the $\mathbb{Z}_2$ topological insulators (TIs), also commonly known as the QSH insulators, which endorse helical edge states protected by the TRS~\cite{KaneII2005,Qi2011}. In contrast to the Chern insulators, these QSH insulators, characterized by a $\mathbb{Z}_2$ invariant (which takes values $0$ or $1$ denoting equivalently \text{``$\mathbb{Z}_2$-even''} (trivial) or \text{``$\mathbb{Z}_2$-odd''} (topological) insulators) demonstrate a zero (charge) Hall current, but carry a quantized spin Hall current owing to two counter-propagating (left and right mover) edge channels for two opposite spins per edge. Moreover, the role of spin-orbit coupling (SOC) that preserves TRS becomes important for a QSH insulator. As TRS is preserved, one can associate a $\mathbb{Z}_2$ invariant with the Kramers pairs that are formed by the occupied states at the Dirac points across the Brillouin zone (BZ). A prototypical model for such $\mathbb{Z}_2$ insulators was proposed by Kane and Mele~\cite{Kane2005} in their seminal paper, where they have considered equal and opposite Haldane flux for the up and down spins of the particles, and thus this spinful version of the Haldane model restores the TRS in the system. In accordance with the TRS, the spinful bands add up to a net zero Chern number (contributing zero charge Hall current), with them being equal and opposite for up ($C_{\uparrow}$) and down-spin ($C_{\downarrow}$) bands. However, a spin dependent Chern number, namely $C_{\sigma}$ (or the difference between $C_{\uparrow}$ and $C_{\downarrow}$) may act as an effective topological invariant giving rise to a finite spin Hall conductivity. Nevertheless, for a more realistic $\mathbb{Z}_2$ TI, this picture fails upon inclusion of an extrinsic SOC, for example, the Rashba SOC (RSOC). Although it protects the TRS, it mixes the spins, destroying the conservation of the $z$ component of the spin. Therefore, $\mathbb{Z}_2$ invariant becomes necessary for such mixed spin systems to encode the topological characterization.

\begin{figure}
\includegraphics[width=0.75\linewidth]{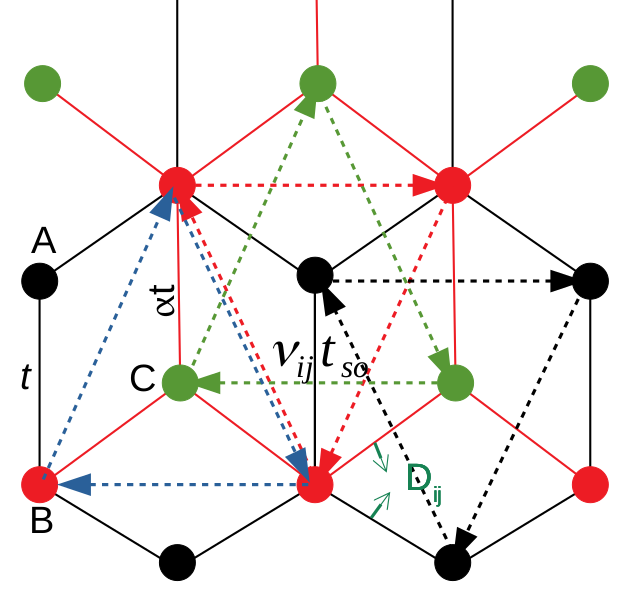}
\caption{The schematic diagram of an $\alpha$-$T_{3}$ lattice is displayed, where the black, red, and green circles represent the sublattices A, B, and C, respectively. The NN hopping strength between A and B sublattices (solid black line) is $t$, while it is $t^\prime=\alpha t$ between B and C sublattices (solid red line). The NN A-C hopping is prohibited here. The NNN hopping between A-B-A (dashed black) or B-A-B (dashed blue) is $t_{so}$, while through C, it is $t_{so}^\prime$ between B-C-B (dashed red) and C-B-C (dashed green), where $t_{so}^\prime=\alpha t_{so}$. $\nu_{ij}=-1 (+1)$ denotes the clockwise (anticlockwise) direction of hopping. The unit vector $\hat{D_{ij}}$ pointing perpendicular to the NN bonds represents the direction of RSOC. The coordinates of the NN sites are ${\bf{d_{1}}}=(\sqrt{3}a_0/2,a_0/2)$, ${\bf{d_{2}}}=(-\sqrt{3}a_0/2,a_0/2)$ and ${\bf{d_{3}}}=(0,-a_0)$, while that of the NNN sites are ${\bf{a_{1}}}=(\sqrt{3}a_0/2,3a_0/2)$, ${\bf{a_{2}}}=(-\sqrt{3}a_0/2,3a_0/2)$ and ${\bf{a_{3}}}=(\sqrt{3}a_0,0)$, $a_0$ being the distance between two neighbouring atoms.}
\label{fig:model}
\end{figure}

The low-lying states of the Dirac electrons present in TIs carry an extra pseudospin-$1/2$ degree of freedom (apart from the usual spins), which can be described by the massless Dirac equation. Until recently, the study of the topology is added to the systems having higher pseudospin, say pseudospin-$1$, whose low-energy modes are described by the Dirac-Weyl Hamiltonian. The $T_3$ or the dice lattice~\cite{Sutherland1986,Xu2017,BerciouxII2011,Vigh2013,Wang2011,Dey2020,Mondal2023,Soni2020,Mohanta2023} is one of such systems which facilitates an extension to the bare graphene (pseudospin-$1/2$) and is designed by an additional atom (\textit{viz.} \text{`C'} sublattice) placed at the center of the hexagonal lattice. Experimentally, this variant of graphene is realized by various mechanisms, such as, by the three counter-propagating laser beams in a cold-atom set-up~\cite{Bercioux2009} or by growing a heterostructure consisting of three-layer cubic lattices, namely, SrTiO$_3$/SrIrO$_3$/SrTiO$_3$ in the (111) direction~\cite{Wang2011}. Interestingly, a more generalized variant of the honeycomb lattice, called the $\alpha$-$T_{3}$ lattice~\cite{Raoux2014,Malcolm2015,Illes2015,Wang2021}, $\alpha$ being the nearest-neighbour (NN) hopping strength between the central atom and one of the sublattices, $\text{`A'}$ or $\text{`B'}$ (shown in Fig.\,\ref{fig:model}) is brought to light. The parameter $\alpha$ smoothly controls the change in the Berry phase (a function of $\alpha$) from $\pi$ (graphene) to $0$ (dice) accompanied by a diamagnetic ($\alpha=0$)-paramagnetic ($\alpha=1$) transition at a critical $\alpha_{c}=0.495$~\cite{Raoux2014} interpolating the two extreme limits, namely the graphene ($\alpha=0$) and the dice ($\alpha=1$). Furthermore, there arises a dispersionless flat band (at zero energy) arises between two dispersive valance and the conduction bands. Apart from the plethora of studies revealing the role of Berry phase~\cite{Raoux2014,Illes2015,Dey2018,Iurov2019,SinghI2023}, valley-polarized transport~\cite{Islam2017,Niu2022} that may be applicable to valleytronics, Klein tunneling~\cite{Illes2017}, optical conductivity~\cite{Illes2015,Illes2016,Kovacs2017,Chen2019,ChenLei2019,Han2022}, magnetotransport properties, such as Shubnikov–de Hass oscillation and quantized Hall conductivity~\cite{Biswas2016,Illes2015,WangJJ2020,SinghI2023}, Floquet dynamics~\cite{Dey2019,Tamang2021}, Majorana corner states~\cite{Mohanta2023}, and other topological signatures~\cite{Soni2020,Mondal2023,BerciouxII2011,Wang2021}, there has also been emerging interest in the study of TPT in an $\alpha$-$T_3$ lattice. Just by itself, topology in multiband systems has emerged as a potential field of research which gave birth to exotic phenomena in kagomé~\cite{Tang2011,Liu2012,Trescher2012,Okamoto2022}, Lieb~\cite{Jaworowski2015}, and dice~\cite{Xu2017,Vigh2013,Wang2011,BerciouxII2011} lattices. Of late, TPT has been reported in an $\alpha$-$T_3$ lattice via Floquet mechanism~\cite{Dey2019} and also via tuning $\alpha$ to induce a QSH transition~\cite{Wang2021}. 

On the other hand, higher order TIs (HOTIs) have become an indispensable part of modern condensed matter systems~\cite{Benalcazar2017,Schindler2018,BenalcazarII2017} due to their unconventional features of \text{`nontrivial'} $(d-n)$-dimensional surface/edge for a $d$-dimensional HOTI of order $n$. Precisely, a 2D second order TI (SOTI) manifests zero-dimensional localized corner modes with gapped (1D) edges, while a 3D SOTI exhibits gapless 1D hinge states with (2D) gapped surface states. The usual notion of bulk boundary correspondence (BBC) fails here and the HOTIs discern a refined BBC. In general, HOTIs are considered a subclass of topological crystalline insulators~\cite{Fu2011,Ando2015} where the nontrivial phases are protected by spatial symmetries like space-inversion, rotation, mirror, etc. An immense interest was observed in this field since Benalcazar~\etal~\cite{Benalcazar2017} have proposed the electric multipole insulators and along with Schindler~\etal~\cite{Schindler2018} have reported the chiral and helical HOTIs which have also claimed experimental realization of HOTIs~\cite{SchindlerII2018,Aggarwal2021} in subsequent years.
    
To our belief, all these previous proposals are mostly based on the single-particle (non-interacting) picture. However, the many-body effects like electron-electron (e-e) and electron-phonon (e-p) interactions in conjuring a TPT in an $\alpha$-$T_3$ system have been scarce. Yet, enormous attempts were made in the past to look for nontrivial topological phases driven by e-e interaction~\cite{Mohanta2023,Soni2020,Okamoto2022,Rachel2018,Raghu2008} (and the references therein), the role of e-p interaction in such context was left unexplored. 
In a polar or an ionic crystal, owing to the interaction between the \text{`oscillating'} lattice and the (extra) fermionic impurity, the quasiparticles, commonly known as \text{`polarons'} are generated which can be described as electrons dressed with phonon clouds. They may get trapped (if the coupling is strong) inside the polarization potential created by this boson-fermion interaction or can move throughout the distorted lattice (for a weak coupling). Consequently, the radius of the polarons depends on the e-p coupling strength. Hence, in a tight-binding lattice (where the electron is bound to its own atomic site), the size of the polaron is smaller compared to the lattice constant and is often referred to as a Holstein polaron~\cite{Holstein1959}. In the past, the e-p interaction has been able to deliver fruitful explanations to the origin of  superconductivity~\cite{Tinkham2004,Frohlich1950,Bardeen1951}, charge density wave and Peierls transition~\cite{Maykon2022,Xie2022,Campetella2023,Luo2022,Miao2023,Casebolt2024,Luo2023,Zhang2023}, and more recently, Bose and Fermi polarons in ultracold gases~\cite{Mostaan2023,Schmidt2022,Koschorreck2012,Scazza2017}, phonon-induced Floquet topological phases~\cite{Chaudhary2020,Hubener2018} etc. We must appreciate the efforts that have been conducted in studying the impact of e-p interaction to engineer the topologically nontrivial phases in a Chern insulator~\cite{Cangemi2019,Camacho2019}, graphene nanoribbon~\cite{Calvo2018}, topological superconductors~\cite{Shaozhi2023}, Dirac systems~\cite{Garate2013,Heid2017,Hu2021} and in other $2D$ materials~\cite{Sergi2020,Pimenov2021,Medina2022,Lu2023} in recent times. These studies significantly lead to the possibility of triggering nontrivial phases and the plausible occurrence of a TPT induced by e-p coupling. However, we must specify that the e-p coupling effect on generating a TPT in a multiband system, for example an $\alpha$-$T_3$ lattice is seldom studied, until very recently, Islam~\etal~\cite{Islam2024} have reported the same in an $\alpha$-$T_3$ Haldane-Holstein model.

However, to the best of our knowledge, there is hardly any study which explores the role of e-p coupling on stimulating nontrivial topological phases and the relative first order TPT (FOTPT) in an $\alpha$-$T_{3}$ QSH system. Neither the evidence of an SOTI phase, nor a second order TPT (SOTPT) has been reported so far in an $\alpha$-$T_3$ lattice. Moreover, the stability of such SOTI phase against any many-body effects, such as e-p interaction, has also not been looked into. Therefore, in this study, we are motivated to investigate the polaron physics in a QSH insulator embedded on an $\alpha$-$T_3$ lattice. Specifically, we shall explain the role of the e-p interaction in inducing a FOTPT, and furthermore, a SOTPT in a Rashba coupled QSH (non-trivially gapped) insulator of the Kane-Mele type. We assume a polaronic impurity (of Holstein type) propagates in a pseudospin-$1$ fermionic system, such as, an $\alpha$-$T_3$ QSH lattice and interacts with the lattice vibrations, which may affect the spectral properties of the bulk bands and result in a bulk gap closing transition solely mediated through the e-p coupling. Our primary goal is the following. At first, we shall inspect whether such interaction can induce a FOTPT accompanied by its conventional signatures, such as an abrupt change in $\mathbb{Z}_2$ invariant, and (dis)appearance of conducting helical edge modes. It will be interesting to see how the TRS can be broken , thereby opening a spectral gap in a QSH insulator and how e-p coupling affects this phase. One such technique, which would do this job is to include an in-plane magnetic field~\cite{Ren2020}.  Hence, we shall encounter how an in-plane magnetic field can generate an SOTI phase in an $\alpha$-$T_3$ lattice and examine the stability of the corner modes against e-p interaction which may persuade a SOTPT upon tuning the e-p coupling without harming the essential symmetries of the system. 

The remainder of the paper is structured as follows. In Sec.~\ref{textFOTI}, we describe the FOTPT via e-p coupling where the description of our system and the corresponding model Hamiltonian is given in Sec.~\ref{textmodel} formulated under the framework of the Kane-Mele model with the Rashba coupling modified by a Holstein term accounting for the e-p coupling. The polaron formation in our system is employed in Sec.~\ref{textLFT} through the Lang-Firsov technique, which works well in the anti-adiabatic (high-frequency) limit and subsequently, a momentum space version of the effective Hamiltonian (obtained in Appendix.~\ref{text kspace Hamiltonian}) is used to compute spectral properties and the relevant topological quantities.   
Sec.~\ref{textbulk} deals with the numerical analysis of the bulk gap closing transitions indicating a plausible occurrence of TPT. Hence, Sec.~\ref{textZ2} and Sec.~\ref{textedge} respectively describe the formulation of the $\mathbb{Z}_2$ invariant and the edge states with their variations in different regimes of e-p coupling therein, which confirm a TPT driven by polarons. Furthermore, a phase boundary is presented in Sec.~\ref{textphase} that summarizes our results for the FOTPT interpolating between graphene and a dice lattice. In Sec.~\ref{textSOTI} we propose the evidence of an SOTI phase induced by an in-plane magnetic field and discuss elaborately how e-p coupling can generate a SOTPT for different $\alpha$ values. Finally, we conclude our findings in Sec. \ref{Sec:summary}.   

\section{Quantum spin Hall transition via e-p coupling}\label{textFOTI}

In this section we discuss the FOTPT mediated through e-p coupling in a pseudospin-$1$ fermionic system on an $\alpha$-$T_{3}$ Kane-Mele (KM) QSH insulator. We schematically represent an $\alpha$-$T_{3}$ lattice in Fig.\,\ref{fig:model}. A typical hexagonal unit cell of an $\alpha$-$T_{3}$ lattice constitutes A (black) and B (red) lattice sites constructing a regular honeycomb lattice with NN hopping strength $t$ between them and an additional C (green) atoms placed at the center of each hexagon connecting to only B atoms (hopping between C and A atom is forbidden) via a hopping strength $\alpha t$ ($\alpha\in [0:1]$). Therefore, the two limiting cases of our study are the results for graphene $(\alpha=0)$ and dice $(\alpha=1)$ lattices. We introduce our model Hamiltonian as follows.

\subsection{Rashba coupled Kane-Mele-Holstein Hamiltonian in an $\alpha$-$T_{3}$ lattice}\label{textmodel}
In order to study the effects of the e-p coupling on an $\alpha$-$T_{3}$ QSH insulator in the presence of a Semenoff mass and the RSOC we model our system under the framework of a tight-binding Kane-Mele-Holstein Hamiltonian, which can be written as $\mathcal{H}=\mathcal{H}_{\text{KM}}+\mathcal{H}_\text{R}+\mathcal{H}_{\mathcal{M}}+\mathcal{H}_{\text{ep}}$ with 

\begin{subequations}
\begin{eqnarray}
\mathcal{H}_{\text{KM}}&=&-t\sum_{\langle i,j\rangle\sigma}c^{\dagger}_{i\sigma}c_{j\sigma}-\alpha t\sum_{\langle i,k \rangle\sigma}c^{\dagger}_{i\sigma}c_{k\sigma}\nonumber\\
&&-\frac{it_{so}}{3\sqrt{3}}\sum_{\langle\langle i,j\rangle\rangle\sigma\sigma^\prime}\nu_{ij}c^{\dagger}_{i\sigma}\sigma_z c_{j\sigma^\prime}\nonumber\\
&&-\frac{i\alpha t_{so}}{3\sqrt{3}}\sum_{\langle\langle i,k\rangle\rangle\sigma\sigma^\prime}\nu_{ik}c^{\dagger}_{i\sigma}\sigma_z c_{k\sigma^\prime},
\label{Ham:KM}
\end{eqnarray}
\begin{eqnarray}
\mathcal{H}_\text{R}&=&-i\beta_R\sum_{\langle i,j\rangle\sigma\sigma^\prime}c^{\dagger}_{i\sigma}(\hat{D}_{ij}.\vec{\tau})_{\sigma\sigma^\prime}c_{j\sigma^\prime}\nonumber\\
&&-i\alpha\beta_R\sum_{\langle i,k\rangle\sigma\sigma^\prime}c^{\dagger}_{i\sigma}(\hat{D}_{ik}.\vec{\tau})_{\sigma\sigma^\prime}c_{k\sigma^\prime},
\label{Ham:R}
\end{eqnarray}
\begin{eqnarray}
\mathcal{H}_{\mathcal{M}}&=&\mathcal{M}\sum_{i\sigma}c^{\dagger}_{i\sigma}S_{z}c_{i\sigma}
\label{Ham:M}
\end{eqnarray}
\begin{eqnarray}
\mathcal{H}_{\text{ep}}&=&\hbar\omega_{0}\biggl[\sum_{i}\biggl(b^{\dagger}_{i}b_{i}+\frac{1}{2}\biggr)+g\sum_{i\sigma}c^{\dagger}_{i\sigma}c_{i\sigma}(b^{\dagger}_{i}+b_{i})\biggr],
\label{Ham:int}
\end{eqnarray}
\end{subequations}
where $\mathcal{H}_{\text{KM}}$, $\mathcal{H}_{\text{R}}$, $\mathcal{H}_{\mathcal{M}}$ and $\mathcal{H}_{\text{ep}}$, respectively represent the Hamiltonian for the bare Kane-Mele model, RSOC, Semenoff mass and e-p interaction present in the system. 
$c^{\dagger}_{i\sigma} (c_{i\sigma})$ denotes the creation (annihilation) operator for electrons corresponding to A, B, and C sublattice with $i$, $j$, and $k$ indices, respectively and $\sigma_z$ is the $z$-component of the Pauli spin matrix, $\sigma$ and $\sigma^\prime$ are the up ($\uparrow$) and down ($\downarrow$) spin indices, respectively.
The first term of Eq.~\eqref{Ham:KM} represented by a single angular bracket, $\langle ... \rangle$ denotes the NN hopping between the A and B sites with hopping amplitude $t$, while due to the presence of C atoms in a typical $\alpha$-$T_{3}$ lattice, the same between the B and C sites is represented with a modified amplitude $t^\prime=\alpha t$, expressed in the second term. The third term in Eq.~\eqref{Ham:KM} represents the next nearest-neighbour (NNN) hopping (denoted by the double angular bracket $\langle\langle... \rangle\rangle$) between A-B-A or B-A-B (hopping A-A via B and B-B via A) with an amplitude $t_{so}$ which arises due to the intrinsic SOC, proposed by Kane and Mele for graphene~\cite{Kane2005}. The B-C-B and C-B-C NNN hoppings with a different hopping strength $t_{so}^\prime=\alpha t_{so}$ are represented by the fourth term of Eq.~\eqref{Ham:KM}. $\nu_{ij}$ or $\nu_{ik}=-1 (+1)$ denotes the clockwise (anticlockwise) hop encoding a difference in a relative sign for the $\uparrow$ and $\downarrow$-spins of the electrons which traverse from site $i$ to NNN $j$ or $k$. Eq.~\eqref{Ham:R} displays the Hamiltonian for the extrinsic RSOC originated due to the breaking of spatial inversion symmetry (however, preserves the TRS) which can be represented by NN hopping between the A and B sites with a strength $\beta_R$ and that between the A and C sites with a strength $\beta_R^\prime=\alpha \beta_R$. $\vec{\tau}=(\tau_x,\tau_y,\tau_z)$ is the Pauli matrix vector and $\hat{D_{ij}}$ ($\hat{D_{ik}}$) is the unit vector along the product $\vec{E_{ij}}\times\vec{r_{ij}}$ ($\vec{E_{ik}}\times\vec{r_{ik}}$), $\vec{E_{ij}}$ ($\vec{E_{ik}}$) and $\vec{r_{ij}}$ ($\vec{r_{ik}}$) being the external electric field and the displacement for the bond $ij$ ($ik$), respectively. It can be easily noted that the RSOC term breaks the mirror symmetry $z\rightarrow-z$ which consequently destroys the spin conservation along the $z$-direction. Further, the inclusion of the Semenoff mass ($\mathcal{M}$) breaks the sublattice (inversion) symmetry, which is written in Eq.~\eqref{Ham:M}, where $S_{z}$ is the $z$-component of the pseudospin-1 matrix. The effect of e-p coupling is represented by the Holstein Hamiltonian written in Eq. \eqref{Ham:int}, where the first term signifies the total phononic onsite (at site $i$) energy in terms of the creation (annihilation) operators, $b^{\dagger}_{i} (b_{i})$ of the longitudinal optical (LO) phonons which oscillate with a dispersionless LO frequency, $\omega_{0}$ and interact with the electrons with a coupling strength $g$, denoted by the second term of Eq.~\eqref{Ham:int}. To obtain an effective electronic Hamiltonian that characterizes the bulk and edge spectra and the relevant topological invariant of our system, we proceed as follows.      
\subsection{Effective electronic Hamiltonian via Lang-Firsov transformation}\label{textLFT}
In a polar crystal, the propagating \text{`extra'} charge carrier distorts the lattice structure which can be formulated by the interaction between the vibrating phonons (bosons) and the charge carrier (fermions). Consequently, through the process of emission and absorption of virtual phonons by the electrons at $T=0$, the quasiparticles, called polarons are formed which are considered as the charge carrier dressed with phonon clouds. Depending on the strength of the interaction, the electrons can be self-trapped by the polarization potential created by the lattice distortion or can move throughout the lattice. For a tight-binding (narrow-band systems) lattice, the size of the polaron is typically less than the lattice constant (not spread over several lattice sites), which allows us to consider only the onsite e-p interaction effect and neglect the contributions from the neighbours, thereby treating them to be weak enough. To investigate the effects of this (small) \text{`Holstein polaron'} on the spectral properties, we first decouple the electron and phonon degrees of freedom by employing the well-known Lang-Firsov transformation (LFT). This is a coherent state transformation of a displaced harmonic oscillator that captures the polaron physics adequately well in the high-frequency (anti-adiabatic) regime, meaning the LO frequency of the phonons is much larger than the other parameters of the system, namely, when $\omega_{0}$ to be much greater than $t,t^\prime, t_{so}, t_{so}^\prime, \beta_R, \beta_R^\prime, \mathcal{M}$ and $g$. The LFT with $\mathcal{S}$ being the generator of the transformation~\cite{Lang1963} can be expressed as
\begin{equation}
    \tilde{\mathcal{H}}=e^{\mathcal{S}}\mathcal{H}e^{-\mathcal{S}},~~~~ \mathcal{S}=g\sum_{i\sigma}c^{\dagger}_{i\sigma}c_{i\sigma}(b^{\dagger}_{i}-b_{i}),
\label{LFT}
\end{equation}
which transforms the total Hamiltonian as 
\begin{widetext}
\begin{eqnarray}
\tilde{\mathcal{H}}&=&-t\biggl[\sum_{\langle i,j\rangle\sigma}c^{\dagger}_{i\sigma}c_{j\sigma}e^{[X_{i}-X_{j}]}+\alpha \sum_{\langle i,k \rangle\sigma}c^{\dagger}_{i\sigma}c_{k\sigma}e^{[X_{i}-X_{k}]}\biggr]-\frac{t_{so}}{3\sqrt{3}}\biggl[\sum_{\langle\langle i,j\rangle\rangle\sigma\sigma^\prime}\nu_{ij}c^{\dagger}_{i\sigma}\sigma_z c_{j\sigma^\prime}e^{[X_{i}-X_{j}]}\nonumber\\
&&+\alpha\sum_{\langle\langle j,k\rangle\rangle\sigma\sigma^\prime}\nu_{ik}c^{\dagger}_{i\sigma}\sigma_z c_{k\sigma^\prime}e^{[X_{i}-X_{k}]}\biggr]-i\beta_R\biggl[\sum_{\langle i,j\rangle\sigma\sigma^\prime}c^{\dagger}_{i\sigma}(\hat{D}_{ij}.\vec{\tau})_{\sigma\sigma^\prime}c_{j\sigma^\prime}e^{[X_{i}-X_{j}]}+\alpha\sum_{\langle i,k\rangle\sigma\sigma^\prime}c^{\dagger}_{i\sigma}(\hat{D}_{ik}.\vec{\tau})_{\sigma\sigma^\prime}c_{k\sigma^\prime}e^{[X_{i}-X_{k}]}\biggr]\nonumber\\
&&+\sum_{i\sigma}c^{\dagger}_{i\sigma}(MS_{z}-g^2 \hbar\omega_{0}I_{3})c_{i\sigma}+\hbar\omega_{0}\sum_{i}b^{\dagger}_{i}b_{i},
\label{Ham: mod model}    
\end{eqnarray}
\end{widetext}
where the $X$-terms in the exponent contain the phonon operators at the $i$-th site as $X_{i}=g(b^{\dagger}_{i}-b_{i})$ and $I_{3}$ is a $3\times3$ identity matrix. In the last step, we neglect the constant \text{`1/2'} factor in the onsite phonon energy term.  

To obtain the effective electronic Hamiltonian eliminating phonon modes from our system, we perform a zero-phonon averaging at $T=0$ which results into
\begin{widetext}
\begin{eqnarray}
\tilde{\mathcal{H}}_{\text{eff}}&=&\langle 0\vert \tilde{\mathcal{H}}\vert 0\rangle=-\tilde{t}\biggl[\sum_{\langle i,j\rangle\sigma}c^{\dagger}_{i\sigma}c_{j\sigma}+\alpha \sum_{\langle i,k \rangle\sigma}c^{\dagger}_{i\sigma}c_{k\sigma}\biggr]-\frac{\tilde{t}_{so}}{3\sqrt{3}}\biggl[\sum_{\langle\langle i,j\rangle\rangle\sigma\sigma^\prime}\nu_{ij}c^{\dagger}_{i\sigma}\sigma_z c_{j\sigma^\prime}+\alpha\sum_{\langle\langle j,k\rangle\rangle\sigma\sigma^\prime}\nu_{ik}c^{\dagger}_{i\sigma}\sigma_z c_{k\sigma^\prime}\biggr]\nonumber\\
&&-i\tilde{\beta}_R\biggl[\sum_{\langle i,j\rangle\sigma\sigma^\prime}c^{\dagger}_{i\sigma}(\hat{D}_{ij}.\vec{\tau})_{\sigma\sigma^\prime}c_{j\sigma^\prime}+\alpha\sum_{\langle i,k\rangle\sigma\sigma^\prime}c^{\dagger}_{i\sigma}(\hat{D}_{ik}.\vec{\tau})_{\sigma\sigma^\prime}c_{k\sigma^\prime}\biggr]+\sum_{i\sigma}c^{\dagger}_{i\sigma}(MS_{z}-g^2 \hbar\omega_{0}I_{3})c_{i\sigma},
\label{Ham: eff model}
\end{eqnarray}
\end{widetext}
where the hopping amplitudes are renormalized by the Holstein reduction factor, namely $\langle 0\vert e^{[X_{i}-X_{j}]}\vert 0\rangle = e^{-g^2}$ as 
\begin{eqnarray}
\tilde{t}=te^{-g^2},~~~~~\tilde{t}_{so}=t_{so} e^{-g^2},~~~~~\tilde{\beta}_R=\beta_R e^{-g^2}.
\label{Holstein amp}
\end{eqnarray}
The last term of Eq.~\eqref{Ham: mod model} drops out due to zero-phonon averaging. Within this anti-adiabatic ($\omega_{0}\gg t,t^\prime, t_{so}, t_{so}^\prime, \beta_R, \beta_R^\prime, \mathcal{M}$, and $g$) decoupling scheme, the electronic and phonon modes do not affect each other any longer because the fast LO phonons possess larger energy gap between different phononic excitations than the electronic hopping strengths, so that the phonons easily follow the electronic motion without altering their distributions.  
It is evident from Eq.~\eqref{Ham: eff model} that the phonon degrees of freedom are removed from the system and the polaronic signatures are well inculcated through the Holstein factors (Eq.~\eqref{Holstein amp}) in the modified electronic hopping strengths, $\tilde{t}$, $\tilde{t}_{so}$, and $\tilde{\beta}_R$ which gives rise to the band narrowing effect and also through the polaron shift energy $-g^2 \hbar\omega_{0}$ (last term of Eq.~\eqref{Ham: eff model}). As these polaronic factors significantly influence the bands, one can tune the band topology accompanied by a bulk gap closing transition via changing the e-p interaction strength.      


\subsection{Bulk spectral properties}\label{textbulk} 
\begin{figure}
\includegraphics[width=1.025\linewidth]{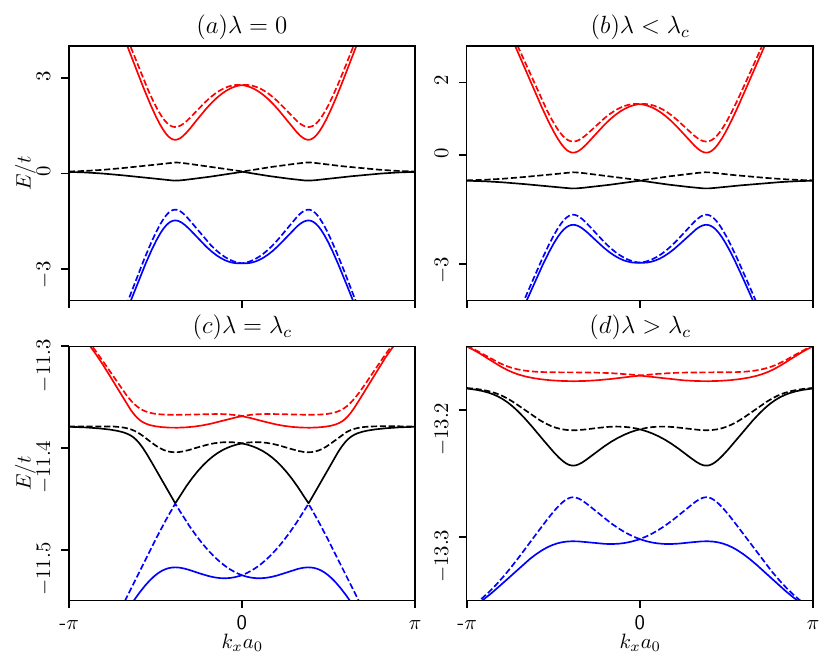}
\caption{The bulk band structures with energy $E$ (in the units of $t$) are shown as a function of dimensionless momenta, $k_x$ (multiplied by $a_0$) at $k_y=2\pi/3a_0$ for $\alpha=0.2$. The red, black, and blue colours represent the CB, the FB and the VBs, respectively. In $(a)$ and $(b)$ the dispersion is plotted in the $0\leq\lambda<\lambda_c$ regime at $\lambda=0$ and $\lambda=0.5$, respectively, where the bulk is gapped. $(c)$ The plot is shown at the critical $\lambda=\lambda_c=1.95$ where the bulk gap closing between the FB and the VB occurs. $(d)$ The same is shown in the $\lambda>\lambda_c$ regime at $\lambda=2.2$ where the spectrum is again gapped. The parameters are taken as $t_{so}=0.1t$, $\mathcal{M}=0.02t$ and $\beta_R=0.02t$.}
\label{fig:bulk_p2}
\end{figure}
\begin{figure}
\includegraphics[width=1.025\linewidth]{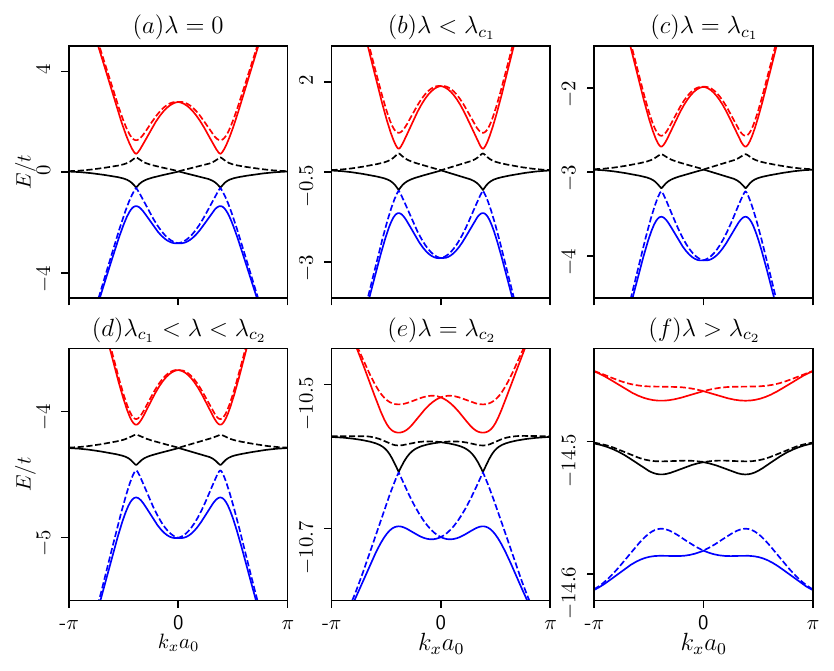}
\caption{The bulk band structures with energy $E$ (in the units of $t$) are shown as a function of dimensionless momenta, $k_x$ (multiplied by $a_0$) at $k_y=2\pi/3a_0$ for $\alpha=0.52$. The red, black, and blue colours represent the CB, the FB and the VBs, respectively. In $(a)$ and $(b)$ the dispersion is plotted in the $0\leq\lambda<\lambda_c$ regime at $\lambda=0$ and $\lambda=0.5$, respectively, where the bulk gap is zero. $(c)$ The plot is shown at the critical $\lambda=\lambda_{c_1}=1.00$ where the bulk gap between the FB and the VB is seen. $(d)$ The same is shown in the $\lambda_{c_1}<\lambda<\lambda_{c_2}$ regime at $\lambda=1.2$ where the spectrum is sufficiently gapped. $(e)$ The bulk closing transition occurs at $\lambda=\lambda_{c_1}=1.88$. $(f)$ Again the spectrum becomes gapped in the $\lambda>\lambda_{c_2}$ regime at $\lambda=2.2$. The parameters are taken as $t_{so}=0.1t$, $\mathcal{M}=0.02t$ and $\beta_R=0.02t$.}
\label{fig:bulk_p52}
\end{figure}
\begin{figure}
\includegraphics[width=1.025\linewidth]{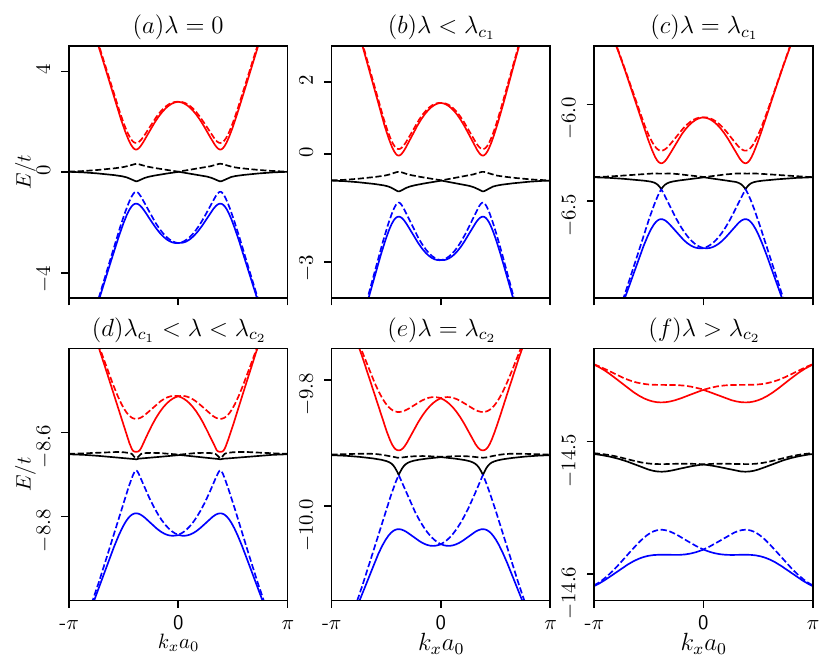}
\caption{The bulk band structures with energy $E$ (in the units of $t$) are shown as a function of dimensionless momenta, $k_x$ (multiplied by $a_0$) at $k_y=2\pi/3a_0$ for $\alpha=0.7$. The red, black, and blue colours represent the CB, the FB and the VBs, respectively. In $(a)$ and $(b)$ the dispersion is plotted in the $0\leq\lambda<\lambda_c$ regime at $\lambda=0$ and $\lambda=0.5$, respectively, where the bulk is gapped. $(c)$ The plot is shown at the critical $\lambda=\lambda_{c_1}=1.46$ where the bulk gap between the FB and the VB is closed. $(d)$ The same is shown in the $\lambda_{c_1}<\lambda<\lambda_{c_2}$ regime at $\lambda=1.70$ where the spectrum is sufficiently gapped. $(e)$ Again bulk closing transition occurs at $\lambda=\lambda_{c_1}=1.82$. $(f)$ The spectrum becomes gapped in the $\lambda>\lambda_{c_2}$ regime at $\lambda=2.2$. The parameters are taken as $t_{so}=0.1t$, $\mathcal{M}=0.02t$ and $\beta_R=0.02t$.}
\label{fig:bulk_p7}
\end{figure}

To study the spectral properties of the QSH $\alpha$-$T_3$ lattice, we Fourier transform Eq.~\ref{Ham: eff model} to the momentum ($\bf k$) space. The details of the calculations are presented in Appendix~\ref{text kspace Hamiltonian}. Subsequent discussions simplify that we shall divide the Eq.~\eqref{Ham:momentum space} by the NN hopping strength, $t$ (all the parameters will be renormalized in the units of $t$) and use the same as our model Hamiltonian for the rest of the paper. Therefore, we introduce a dimensionless parameter (renormalized e-p coupling strength) as $\lambda=\frac{g^2\omega_0}{t}$ which we henceforth call the e-p coupling constant for our system.

For convenience, we consider $a_0=\hbar=1$. To perform the numerical analysis for the FOTPT in a Kane-Mele $\alpha$-$T_3$ lattice in the presence of RSOC, we fix the hopping strengths as $t=-2.8$ eV, $t_{so}=0.1t$, $\beta_R=0.02t$ and the Semenoff mass as $\mathcal{M}=0.02t$. The rationale behind choosing these values is inspired by the corresponding quantities in graphene. We also set the LO frequency of the phonons, $\omega_0$ as $\omega_0=3t\gg t,t^\prime, t_{so}, t_{so}^\prime$, $\beta_R, \beta_R^\prime$, and $\mathcal{M}$ (for anti-adiabaticity to be valid).

The distinction of an $\alpha$-$T_3$ lattice from bare graphene ($\alpha=0$) lies in the appearance of FB for $\alpha\neq 0$, thus justifying pseudospin-$1$ fermions. Hence, the role of the FB needs to be ascertained carefully in the context of TPT. A conventional $\alpha$-$T_3$ lattice is identified with a dispersionless flat band (FB) at zero-energy for graphene ($\alpha=0$) and dice ($\alpha=1$) lattices and a distorted FB for $0<\alpha<1$ along with two dispersive bands, namely the conduction (CB) and the valance (VB) band which touch each other at the Dirac cones located at $\textbf{K}=(2\pi/3\sqrt 3a_0,2\pi/3a_0)$ and ${\bf{K^\prime}} = (-2\pi/3\sqrt 3a_0,2\pi/3a_0)$. For our case, due to the presence of the intrinsic SOC (KM) term, the FB becomes more dispersive, but the valley degeneracy is preserved due to the TRS, unlike the $\alpha$-$T_3$-Haldane lattice~\cite{Islam2024}. The distorted FB favours the transport of the system as the band electrons are associated with a non-zero group velocity. 
As the bulk band properties are different for different values of $\alpha$, we segregate the whole range of $\alpha$ ($[0:1]$) into two regimes, namely $(i)$ the lower to intermediate ($0<\alpha\leq 0.5$) and $(ii)$ intermediate to higher ($0.5<\alpha<1.0$) values and present the spectral properties for a particular $\alpha$ from each of the two regimes. The results for lower values of $\alpha$ are close to those of bare graphene ($\alpha=0$), while those for higher $\alpha$ agree with the dice ($\alpha=1$) lattice. Therefore, we encompass the entire range of $\alpha$ interpolating between the graphene and the dice lattice.  

At first, let us consider the bulk spectra for $\alpha=0.2$ which are presented in Fig.\,\ref{fig:bulk_p2}. The spectra are plotted as a function of dimensionless momentum, $k_xa_0$ ($k_y$ is fixed at $\frac{2\pi}{3a_0}$), where the solid and the dotted lines represent those for the up and the down spins, respectively for the CB (red), FB (black) and the VB (blue). One may notice that the dispersions of the bands are of semi-Dirac type, which means, they are linear along $k_y$, while quadratic along the $k_x$ direction. The low energy spectrum of the Hamiltonian \eqref{Ham:momentum space} obeys the Dirac-Weyl dispersion, which can be achieved by linearizing Eq.~\eqref{Ham:momentum space} in the vicinity of the two valleys $\textbf{K}$ and $\bf{K^\prime}$. As the SOC breaks the spin-degeneracy, the bulk bands are split for the $\uparrow$-spins and the $\downarrow$-spins giving rise to six distinct bands (three for each spin orientation). However, we cannot distinguish them as $\uparrow$ or $\downarrow$-spin bands as the spin-polarization is lost due to the presence of RSOC ($\beta_R$). The bulk bands corresponding to the $\uparrow$-spins and the $\downarrow$-spins cross each other only at the $M$ $(0,2\pi/3a)$ point, resulting in a zero spin-splitting.  Moreover, the TRS validates identical behaviour of the bulk bands at the two valleys, namely $\textbf{K}$ and ${\bf{K^\prime}}$. To investigate the different topological phases and the TPT therein mediated via polaron, we must vary the e-p coupling ($\lambda$) and examine possible gap closing transitions at $\lambda=\lambda_c$,  $\lambda_c$ being the e-p coupling at the TPT. The different regimes of $\lambda$ are entitled as $0\leq\lambda<\lambda_c$ (before the transition), $\lambda=\lambda_c$ (at the transition) and $\lambda>\lambda_c$ (beyond the transition). We can clearly see from Fig.\,\ref{fig:bulk_p2}$(a)$ that the bulk bands are gapped at both valleys when $\lambda=0$, which implies that the system for small $\alpha$, for example, $\alpha=0.2$ behaves as an insulator in the absence of e-p coupling. Further, as $\lambda$ increases, the overall spectrum shifts downward (because of the polaronic shift energy $-g^2 \hbar\omega_{0}$), and the FB is no longer fixed at $E=0$ in the $\lambda<\lambda_c$ regime (Fig.\,\ref{fig:bulk_p2}$(b)$). However, the gap in the spectrum still remains intact. As $\lambda$ reaches a critical value, namely $\lambda=\lambda_c=1.95$, we find that the bulk gap between the FB and the VB closes (shown in Fig.\,\ref{fig:bulk_p2}$(c)$) at both the valleys, which indicates a bulk gap closing transition. Beyond, $\lambda=\lambda_c$ (i.,e, in $\lambda>\lambda_c$ regime), there exists a noticeable gap in the bulk spectrum (shown in Fig.\,\ref{fig:bulk_p2}$(d)$). Hence, for the lower values of $\alpha$, the bulk bands are gapped in the $0\leq\lambda<\lambda_c$ and $\lambda>\lambda_c$ regimes, resulting in an insulating behaviour, while at $\lambda=\lambda_c$ they are gapless, implying a semi-metallic ($SM$) point. As our system is modelled by the Kane-Mele Hamiltonian, one of the gapped phases may carry the signature of a QSH insulating phase. To verify whether a certain gapped phase corresponds to a topological one, we calculate the topological invariant, namely the $\mathbb{Z}_2$ invariant, in the subsequent section. Therefore, a FOTPT is possible for an $\alpha$-$T_3$ QSH lattice solely induced by the e-p coupling, which is one of the pivotal points of our study. We must specify that similar findings hold true for the other values of $\alpha$ in the $0<\alpha\leq 0.5$ range (however, not shown here).    
\begin{table}[h!]
    \begin{tabular}{|c|c|c|c|c|c|c|} 
      \hline
      ${\bm{\alpha}}$ & ${\bm{\lambda_{c}}}$ & ${\bm{g_{c}}}$ & ${\bm{\lambda_{c_1}}}$ & ${\bm{\lambda_{c_2}}}$ & ${\bm{g_{c_1}}}$ & ${\bm{g_{c_2}}}$\\
      \hline
      0.1 & 1.97 & 0.81 & - & - & - & -\\
      \hline
      0.2 & 1.95 & 0.81 & - & - & - & -\\
      \hline
      0.3 & 1.92 & 0.80 & - & - & - & -\\
      \hline
      0.4 & 1.90 & 0.81 & - & - & - & -\\
      \hline
      0.5 & 1.90 & 0.79 & - & - & - & -\\
      \hline
      0.52 & - & - & 1.00 & 1.88 & 0.51 & 0.78\\
      \hline
      0.6 & - & - & 1.28 & 1.85 & 0.65 & 0.79\\
      \hline
      0.7 & - & - & 1.46 & 1.82 & 0.69 & 0.78\\
      \hline
      0.8 & - & - & 1.60 & 1.78 & 0.73 & 0.77\\
      \hline
      0.9 & - & - & 1.67 & 1.73 & 0.74 & 0.76\\
      \hline
      \end{tabular}
      \caption{Table of transition points, namely $\lambda_{c}$, corresponding to the bare e-p coupling strengths at the transitions, namely $g_c$, related as $\lambda_c=\frac{g_c^2\omega_0}{t}$, for different $\alpha$ cases. While $g_c$ is always less than $1$, $\lambda_c$ acquires values greater than $1$ due to scaling by $\omega_0/t$.}
      \label{tab:table1}
      \quad
\end{table}

The scenarios become contrasting for intermediate to higher values of $\alpha$ ($0.5<\alpha<1.0$) which are shown in Figs.\,\ref{fig:bulk_p52} and \ref{fig:bulk_p7}. While deriving the band spectra for the intermediate $\alpha$ values, we come across a very interesting observation for $\alpha=0.52$, presented in Fig.\,\ref{fig:bulk_p52}$(a)$. It is indeed a special point at which the bulk band gap is zero even when $\lambda=0$, which corresponds to an $SM$ phase of the $\alpha$-$T_3$ lattice. Earlier studies suggest that a TPT in the vicinity of this point~\cite{Wang2021} exists in the absence of any interaction, supported by a gap closing transition in the bulk spectra of the $\alpha$-$T_3$ lattice. Here, the FB becomes more distorted, as it should be, for larger values of $\alpha$. This $SM$ phase for $\alpha=0.52$ remains unchanged for $\lambda<\lambda_{c_1}$ (Fig.\,\ref{fig:bulk_p52}$(b)$) up to $\lambda=\lambda_{c_1}$, where $\lambda_{c_1}=1$, where we observe a very small gap (may be zoomed in for better visualization) for the first time between the FB and the VB, shown in Fig.\,\ref{fig:bulk_p52}$(c)$. Consequently, in the $\lambda_{c_1}<\lambda<\lambda_{c_2}$ regime (Fig.\,\ref{fig:bulk_p52}$(d)$), a noticeable gap appears in the bulk spectrum which closes again at $\lambda=\lambda_{c_2}=1.88$ (Fig.\,\ref{fig:bulk_p52}$(e)$), beyond which the FB and the VB are again gapped referring it to an insulating state (Fig.\,\ref{fig:bulk_p52}$(f)$). Therefore, in the vicinity of $\alpha=0.52$, the system undergoes an $SM$-insulator-insulator transition as the e-p coupling is enhanced. As pointed out earlier, we shall investigate the topological characterizations of these phases later by computing the $\mathbb{Z}_2$ invariant. 
However, for higher range of $\alpha$ ($0.6\leq\alpha<1.0$), the existence of the $SM$ phase for the $\lambda<\lambda_{c_1}$ regime is not certain. To clarify that, we present the results for $\alpha=0.7$ as a sample case of the higher $\alpha$ regime. Interestingly, for $\alpha=0.7$, a band gap is again seen between the FB and the VB (Fig.\,\ref{fig:bulk_p7}$(a)$) for $\lambda=0$ as well as in the $\lambda<\lambda_{c_1}$ region (Fig.\,\ref{fig:bulk_p7}$(b)$) till $\lambda$ becomes $\lambda_{c_1}=1.46$ (Fig.\,\ref{fig:bulk_p7}$(c)$). Beyond this value the system enters into another gapped phase in the $\lambda_{c_1}<\lambda<\lambda_{c_2}$ (Fig.\,\ref{fig:bulk_p7}$(d)$) and finally remains gapped in the $\lambda>\lambda_{c_2}$ regime (Fig.\,\ref{fig:bulk_p7}$(f)$) accompanied by a gap closing transition at $\lambda=\lambda_{c_2}=1.82$ (Fig.\,\ref{fig:bulk_p7}$(e)$). The findings of $\alpha=0.7$ are also valid for any value of $\alpha$ in range $0.6\leq\alpha< 1.0$, albeit with different $\lambda_{c_1}$ and $\lambda_{c_2}$. It is to be noted that as we sufficiently increase $\alpha$, the difference between $\lambda_{c_1}$ and $\lambda_{c_2}$ reduces, merging into a single transition point, as observed for the lower $\alpha$ ($0<\alpha\leq 0.5$) regime, especially for $0.9<\alpha<1.0$ (close to the dice lattice). 
Hence, for $0.6\leq\alpha<1.0$, the system facilitates a re-entrant mechanism to multiple insulating phases accompanied by two gap closing transitions mediated through e-p coupling, while the lower ($0<\alpha\leq 0.5$) $\alpha$ values show an insulator-insulator transition. The bulk gap closing phenomena can be explained by the two main artefacts of our study: one is the polaronic effects which results in narrowing of bands (Eq.~\eqref{Holstein amp}), and the other can be seen from the matrix elements of $\rho$-block (Eqs.~\eqref{matrho}) which demonstrates the interplay between the three quantities, namely the parameter $\alpha$ (that accounts for the behaviour of the FB), the Semenoff mass $\mathcal{M}$ (that acts as a staggered sublattice potential) and the e-p coupling parameter $\lambda$. Furthermore, these factors affect the FB and the VB and thereby manipulate the gap between them, predominantly for higher values of $\alpha$, which gives rise to a more complex nature of transition in this regime of $\alpha$ (namely, $0.5<\alpha<1.0$). Thus, we may encompass various (topologically different) kinds of nontrivial phases and the relative phase transitions induced by the e-p coupling for all values of $\alpha$ in range $[0:1]$, albeit with different transition points (the values of $\lambda_c$s listed for each $\alpha$ value in Table \ref{tab:table1}). The data contained therein can be broadly divided into two categories: $(i)$ lower $\alpha$ cases with one gap closing point, namely $\lambda_c$ and $(ii)$ intermediate to higher $\alpha$ cases accompanied by two transition points, namely $\lambda_{c_1}$ and $\lambda_{c_2}$. Some of the spectral properties are similar to the ones obtained by some of us earlier, albeit in the context of a Haldane-Holstein model~\cite{Islam2024}.  

The plausible topological invariant to characterize a QSH phase (exhibiting a topological gap in the bulk spectrum) is the $\mathbb{Z}_2$ topological invariant (especially when spins are mixed), which we discuss below. Hence, to confirm the existence of a spin-mixed QSH phase in our system, we consider the evolution of the Wannier charge center (WCC) that determines the $\mathbb{Z}_2$ invariant.    
\subsection{\texorpdfstring{$\mathbb{Z}_2$}{TEXT} invariant: evolution of the Wannier charge center}\label{textZ2}  
In this section, we study how the evolution of the WCC (which is known as the center of charge in a unit cell) in different regimes of $\lambda$ that indicate topologically trivial or non-trivial characteristics for different values of $\alpha$. The WCC can be thought of as the expectation value of the position operator for a basis represented by Wannier functions (WFs). These WFs are a set of orthogonal functions indexed by a lattice position, say $\bf R$, and maximally localized about that point with respect to all relevant spatial dimensions, which can be mathematically expressed as
\begin{eqnarray}
{|W_n(\bf R)\rangle}=\frac{V}{(2\pi)^D}\int_{BZ} d^Dk~e^{-i\bf k.\bf R}|\psi_{n\bf k}\rangle,
\label{WF}
\end{eqnarray}
where $|\psi_{n\bf k}\rangle$ represents the Bloch wavefunction labelled by the band index $n$ and the crystal momentum $\bf k$. $D$ and $V$, respectively, represent the dimensionality of the $k$ space and the volume of the real-space primitive unit cell. On the other hand, a convenient strategy for a 2D QSH may be to construct a \text{``hybrid Wannier functions''} (HWFs), which are localized along one spatial dimension, say along $\hat{x}$ (Wannier like in 1D), and delocalized  along the other directions,namely, $\hat{y}$ and $\hat{z}$ (Bloch like in 2D) which read as
\begin{eqnarray}
{|W_{n}^H(R_x, k_y,k_z)\rangle}=\frac{1}{(2\pi)}\int_{-\pi}^\pi dk_x~e^{-ik_xR_x}|\psi_{n\bf k}\rangle.
\label{WF}
\end{eqnarray}
With the help of the evolution of HWFs around a closed loop in the BZ, the adiabatic, unitary evolution of the occupied Bloch bands can be described. Hence, we can construct the hybrid WCC (HWCC) which can be represented as the expectation value of the position operator, $\hat{\bf X}$, with respect to the HWFs as
\begin{eqnarray}
{\langle x_{n}(k_y,k_z)\rangle}=\langle W_{n}^H(R_x, k_y,k_z)|\hat{\bf X}|W_{n}^H(R_x, k_y,k_z)\rangle.
\label{WF}
\end{eqnarray}
By the modern theory of polarization, the HWCC can be interpreted in terms of the Berry phase, $\phi_n(k_y,k_z)$ as~\cite{Soluyanov2011,Taherinejad2014,Gresch2017}
\begin{equation}
{\langle x_{n}(k_y,k_z)\rangle}=\frac{\phi_n(k_y,k_z)}{2\pi}=\frac{1}{2\pi}\int_0^{2\pi}A_n(k_x,k_y,k_z)dk_x,
\label{WF}
\end{equation}
where $A_n(k_x,k_y,k_z)=-i\langle u_{n\bf k}|\nabla_{\bf k}|u_{n\bf k}\rangle$ denotes the Berry connection which captures the topological signatures of the system and $|u_{n\bf k}\rangle$ refers to the cell-periodic part of the Bloch wavefunction $|\psi_{n\bf k}\rangle$. 

Since our 2D model lies in the $x$-$y$ plane, we can define a general $k$-vector in the momentum space as ${\bf{k}}=k_{\hat{b}_1}\hat{b}_1+k_{\hat{b}_2}\hat{b}_2$ and characterize the $\mathbb{Z}_2$ invariant by employing the hybrid Wannier transformation along $\hat{b}_1$ and studying its evolution as a function of the remaining momentum, that is $k_{\hat{b}_2}$, where $\bf {b_1}$ and $\bf {b_2}$ are the reciprocal lattice vectors and are given as ${\bf {b_1}}=(-\frac{2\pi}{\sqrt{3}a_0},\frac{2\pi}{3a_0})$ and ${\bf {b_2}}=(\frac{2\pi}{\sqrt{3}a_0},\frac{2\pi}{3a_0})$. Therefore, we can now compute the HWCC in the direction $a_1$ as a function of $k_{\hat{b_1}}$, which can be mathematically formulated as $\langle r_{n,\hat{a}_1}(k_{\hat{b}_2})\rangle=\langle W_{n}^H(R_{\hat{a}_1}, k_{\hat{b}_2})|\hat{\bf X}|W_{n}^H(R_{\hat{a}_1}, k_{\hat{b}_2})\rangle$, and assumes a form,
\begin{eqnarray}
{\langle r_{n,\hat{a}_1}(k_{\hat{b}_2})\rangle}=\frac{\phi_n(k_{\hat{b}_2})}{2\pi}=\frac{1}{2\pi}\oint A_n^{\hat{b}_1}(k_{\hat{b}_1},k_{\hat{b}_2})dk_{\hat{b}_1},
\label{WF}
\end{eqnarray}
with $A_n^{\hat{b}_1}(k_{\hat{b}_1},k_{\hat{b}_2})=-i\langle u_{n\bf k}|\frac{\partial}{\partial k_{\hat{b}_1}}|u_{n\bf k}\rangle$. The interpretation of the $\mathbb{Z}_2$ invariant in terms of the HWCC goes as follows. It is defined as the number of individual HWCC crossed by an arbitrary line traversing half the BZ, modulo $2$~\cite{Gresch2017}. If the line intersects an even (odd) number of HWCCs while traversing through half the BZ, the $\mathbb{Z}_2$ invariant is zero (non-zero), assuring a topologically trivial (nontrivial) phase.
\begin{figure}
\includegraphics[width=1.025\linewidth]{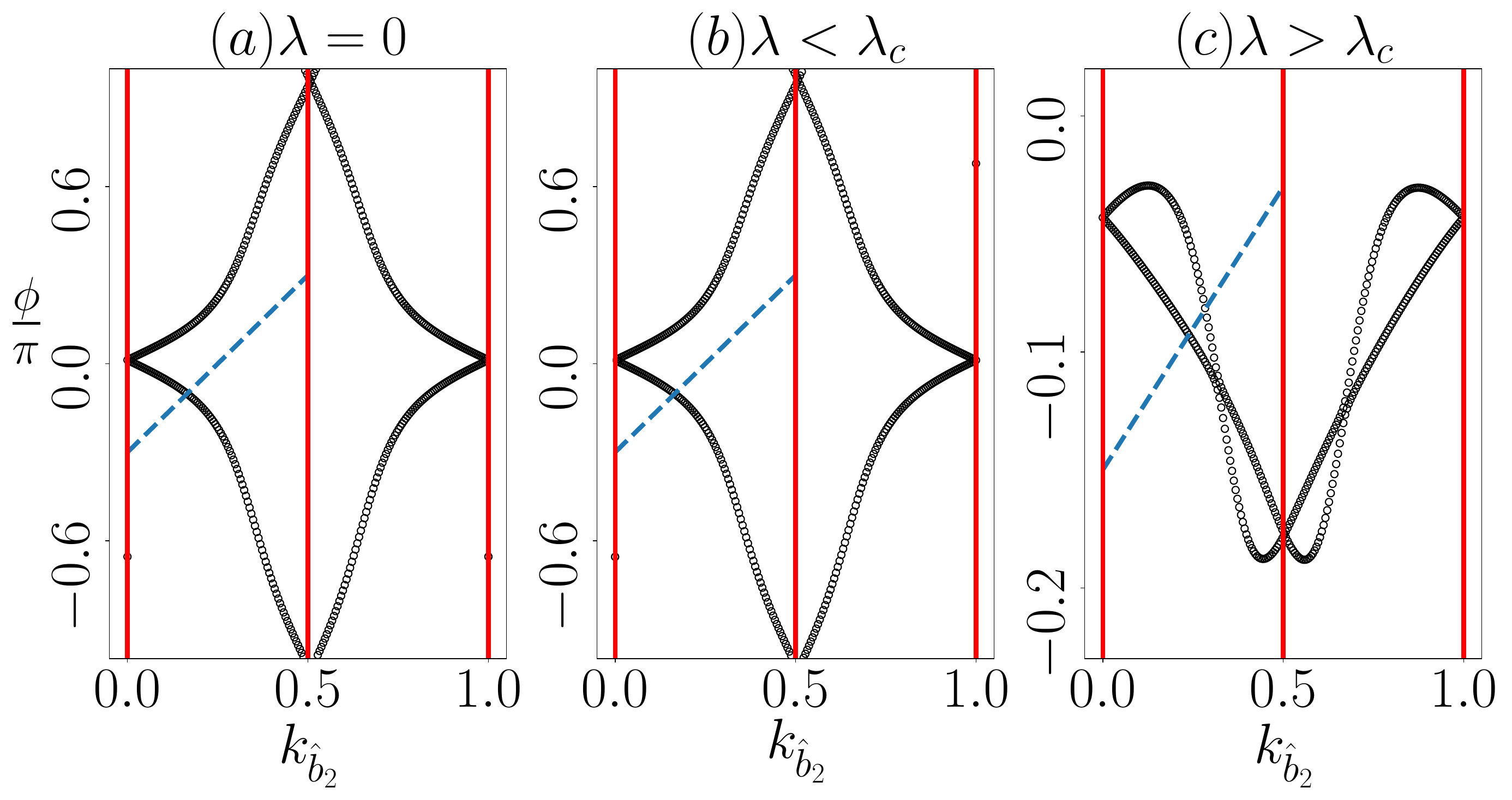}
\caption{The evolution of the HWCCs in terms of the Berry phase $\phi$ (scaled by $\pi$) is shown as a function of dimensionless momenta, $k_{\hat{b}_2}$ for $\alpha=0.2$ in different regimes of $\lambda$, namely $(a)$ $\lambda=0$ and $(b)$ $\lambda<\lambda_c$ ($\lambda=0.5$) where the arbitrary line (dashed blue) cuts through odd number of HWCCs referring to a $\mathbb{Z}_2$-odd topological insulator, while $(c)$ describes the $\lambda>\lambda_c$ regime ($\lambda=2.2$) where the line crosses even number of HWCCs denoting a $\mathbb{Z}_2$-even trivial insulator. The parameters are kept the same as those in Fig.\,\ref{fig:bulk_p2}. The values of $\lambda_{c}$ are mentioned in Table \ref{tab:table1}.}
\label{fig:Z2p2}
\end{figure}
\begin{figure}
\includegraphics[width=1.025\linewidth]{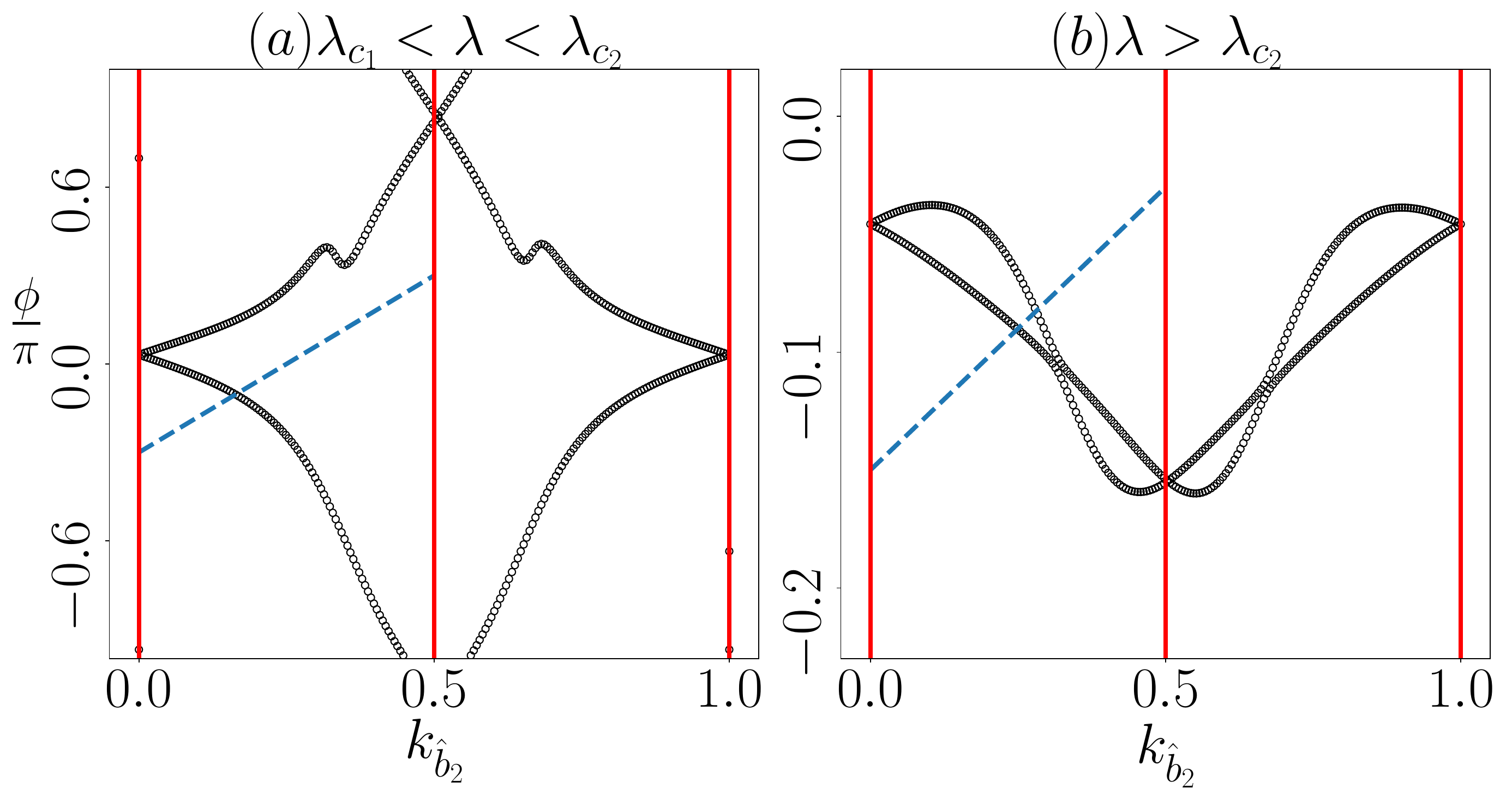}
\caption{The evolution of the HWCCs (in terms of the Berry phase $\phi$) is shown as a function of dimensionless momenta, $k_{\hat{b}_2}$ for $\alpha=0.52$ in different regimes of $\lambda$, namely $(a)$ $\lambda_{c_1}<\lambda<\lambda_{c_2}$ ($\lambda=1.2$) where the arbitrary line (dashed blue) cuts through odd number of HWCCs referring to a $\mathbb{Z}_2$-odd topological insulator, and $(b)$ $\lambda>\lambda_{c_2}$ ($\lambda=2.2$) where the same crosses even number of HWCCs denoting a $\mathbb{Z}_2$-even trivial insulator. The parameters are kept the same as those in Fig.\,\ref{fig:bulk_p52}. The values of $\lambda_{c}$ are mentioned in Table \ref{tab:table1}.}
\label{fig:Z2p52}
\end{figure}
\begin{figure}
\includegraphics[width=1.025\linewidth]{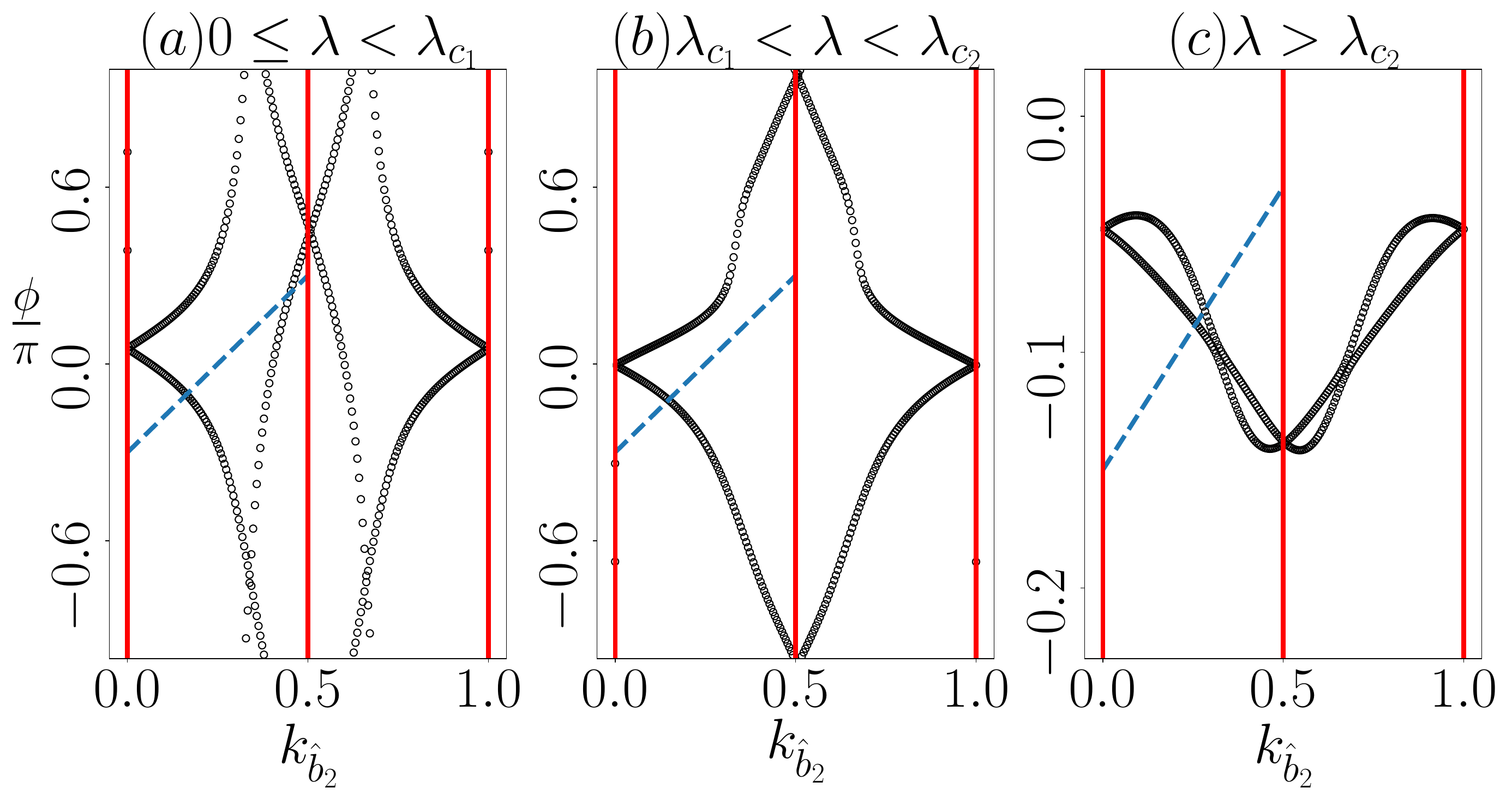}
\caption{The evolution of the HWCCs (in terms of the Berry phase $\phi$) is shown as a function of dimensionless momenta, $k_{\hat{b}_2}$ for $\alpha=0.2$ in different regimes of $\lambda$, namely $(a)$ $0\leq\lambda<\lambda_{c_1}$ ($\lambda=0$) where the arbitrary line (dashed blue) cuts through even number of HWCCs denoting a $\mathbb{Z}_2$-even trivial insulator, $(b)$ $\lambda_{c_1}<\lambda<\lambda_{c_2}$ ($\lambda=1.65$) where it crosses odd number of HWCCs signifying a $\mathbb{Z}_2$-odd topological insulator, and $(c)$ $\lambda>\lambda_c$ ($\lambda=2.2$) where the same intersects even number of HWCCs referring a trivial insulator. The parameters are kept the same as those in Fig.\,\ref{fig:bulk_p7}. The values of $\lambda_{c}$ are mentioned in Table \ref{tab:table1}.}
\label{fig:Z2p7}
\end{figure}

As discussed earlier, the notion of the Chern number fails (being odd under TR) for a TR invariant (TRI) QSH insulator, and the topological properties are guaranteed both by the evolution of WCCs and the surface/edge energy spectra. For a TRS-preserved system, the WCCs obey all the symmetries that are found in the energy bands. Particularly, the existence of the Kramers doublets at the two TRI momenta in the BZ also leads to a double degeneracy in the WCCs. Here, we consider only the first half of the BZ ($[0,\pi/a]$) as the TRS imposes the behaviour of the WCCs and the surface energy bands to be identical to those in the other half of the BZ. To envisage the existence of nontrivial phases followed by a TPT, we thoroughly investigate the evolution of the HWCCs for different regimes of $\lambda$ vis-à-vis the bulk spectra presented in Sec.~\ref{textbulk}.

We present the evolution of the HWCCs for three values of $\alpha$, one in the lower regime ($\alpha=0.2$), one corresponding to the special case at $\alpha=0.52$, and the other in the higher regime ($\alpha=0.7$) of $\alpha$. Consider a 2D $\mathbb{Z}_2$-insulator in the $x$-$y$ plane whose evolution of the HWCCs as a function of the momentum, $k_{\hat{b}_2}$ is displayed in Fig.\,\ref{fig:Z2p2}, Fig.\,\ref{fig:Z2p52}, and Fig.\,\ref{fig:Z2p7} for $\alpha=0.2$, $\alpha=0.52$, and $\alpha=0.7$, respectively. It is clearly visible from Figs.\,\ref{fig:Z2p2}$(a)$ and $(b)$ that for $\alpha=0.2$, the arbitrary (dashed blue) line in the half-BZ cuts through odd (one) number of HWCCs in the $0\leq\lambda<\lambda_c$ regime, which makes the $\mathbb{Z}_2$ invariant non-zero. For a TRS-preserved system, the non-zero $\mathbb{Z}_2$ invariant demonstrates that an HWCC switches its TRI partner as $k_{\hat{b}_2}$ evolves from $0$ to $\pi/a$, and consequently the TR polarization changes sign. Alternatively, it describes that one HWCC jumps from one unit cell to the other (often referred to as TR \textit{ polarization pumping}), signifying a finite conduction on the surface or the edges, which is a typical signature of a $\mathbb{Z}_2$-odd topological (QSH) insulator. However, for $\lambda>\lambda_c$ (Fig.\,\ref{fig:Z2p2}$(c)$), the line crosses even (two) number of HWCCs implying no conduction along the edges, which proclaims a trivial ($\mathbb{Z}_2$-even) insulator. These findings are a direct consequence of the bulk band spectra found in (Fig.\,\ref{fig:bulk_p2}). Although we only show the results for $\alpha=0.2$, the above observations hold for any value in range $0<\alpha\leq 0.5$. Therefore, it is confirmed that in the lower regime of $\alpha$, the system undergoes a topological-trivial transition induced by the e-p coupling.
It is worth mentioning that studying the variations of the $\mathbb{Z}_2$ invariant as a function of $\lambda$ for $\alpha=0.52$ needs special attention as the bulk spectrum is gapless in the $\lambda<\lambda_{c_1}$ regime even at $\lambda=0$ (unlike the other values of $\alpha$ where it is gapped). Therefore, for $\alpha=0.52$, the $\mathbb{Z}_2$ invariant is ill-defined in the $\lambda<\lambda_{c_1}$ regime (including at $\lambda=0$) which allows us to compute the $\mathbb{Z}_2$ invariant only for $\lambda_{c_1}<\lambda<\lambda_{c_2}$ and $\lambda>\lambda_{c_2}$ where the bulk is fairly gapped (Figs.\,\ref{fig:bulk_p52}$(d)$ and $(f)$) denoting the two insulating regions. Fig.\,\ref{fig:Z2p52}$(a)$ reveals that the arbitrary line passes though odd (one) number of HWCCs in the $\lambda_{c_1}<\lambda<\lambda_{c_2}$ regime which demonstrates a topologically nontrivial insulating phase, while we notice an even (two) number of crossings in the HWCCs in the $\lambda>\lambda_{c_2}$ regime (Fig.\,\ref{fig:Z2p52}$(b)$) which confirms the insulating phase in this regime of $\lambda$ to be a trivial one. 

As suggested by the variations of the bulk spectra, the scenario becomes more interesting in the higher regime ($0.6\leq\alpha<1.0$) of $\alpha$ where the system accompanies two bulk gap closing transitions at $\lambda_{c_1}$ and $\lambda_{c_2}$ (for $\lambda_c$ values see Table \ref{tab:table1}), associated with three gapped phases for $0\leq\lambda<\lambda_{c_1}$, $\lambda_{c_1}<\lambda<\lambda_{c_2}$, and $\lambda>\lambda_{c_2}$. We evaluate the results for $\alpha=0.7$ as a representative case for the $0.6\leq\alpha<1.0$ regime whose bulk spectra are discussed in Fig.\,\ref{fig:bulk_p7}. Fig.\,\ref{fig:Z2p7}$(a)$ is plotted for $\alpha=0.7$ in the regime $0\leq\lambda<\lambda_{c_1}$ (where the bulk spectra is widely gapped and are represented in Figs.\,\ref{fig:bulk_p7}$(a)$ and $(b)$). They display an even (two) number of crossings in the HWCCs refereeing to a non-topological phase, while the gapped phase (see the bulk spectrum in Fig.\,\ref{fig:bulk_p7}$(d)$) in the $\lambda_{c_1}<\lambda<\lambda_{c_2}$ regime denotes a topological one (accompanied by finite conduction along the edges) as the line intersects through only one HWCC, shown in Fig.\,\ref{fig:Z2p7}$(b)$. Hence, for higher $\alpha$ ($0.6\leq\alpha<1.0$) cases, nontrivial topological phases emerge in the $\lambda_{c_1}<\lambda<\lambda_{c_2}$ regime solely by the e-p coupling as we moderately tune $\lambda$ beyond $\lambda=\lambda_{c_1}$ (till $\lambda=\lambda_{c_2}$), which are otherwise absent in range $0\leq\lambda<\lambda_{c_1}$. As expected, the regime $\lambda>\lambda_{c_2}$ describes a trivial phase (Fig.\,\ref{fig:Z2p7}$(c)$) for the higher $\alpha$ cases. All the other $\alpha$ cases in range $0.6\leq\alpha<1.0$ show similar results as that for $\alpha=0.7$, albeit with different $\lambda_{c_1}$ and $\lambda_{c_2}$ (not shown here). Thus, the evolution of the HWCCs ascertains that the system for higher values of $\alpha$ ($0.6\leq\alpha<1.0$) facilitates trivial-topological-trivial TPT upon suitably varying the strength of the e-p interaction, $\lambda$.

In order to support the evidence of a TPT mediated through polaronic interactions, we also study the variations of the edge spectra of an $\alpha$-$T_3$ semi-infinite zigzag ribbon as a function of $\lambda$, which we discuss next. We expect the $\mathbb{Z}_2$-odd topological phases (accompanied by an odd number of HWCCs crossings) to exhibit a pair of helical QSH edge modes per edge of the 2D $\alpha$-$T_3$ lattice, with them vanishing for the topologically trivial phases.
\subsection{Edge spectral properties}\label{textedge} 
\begin{figure}
\includegraphics[width=\columnwidth]{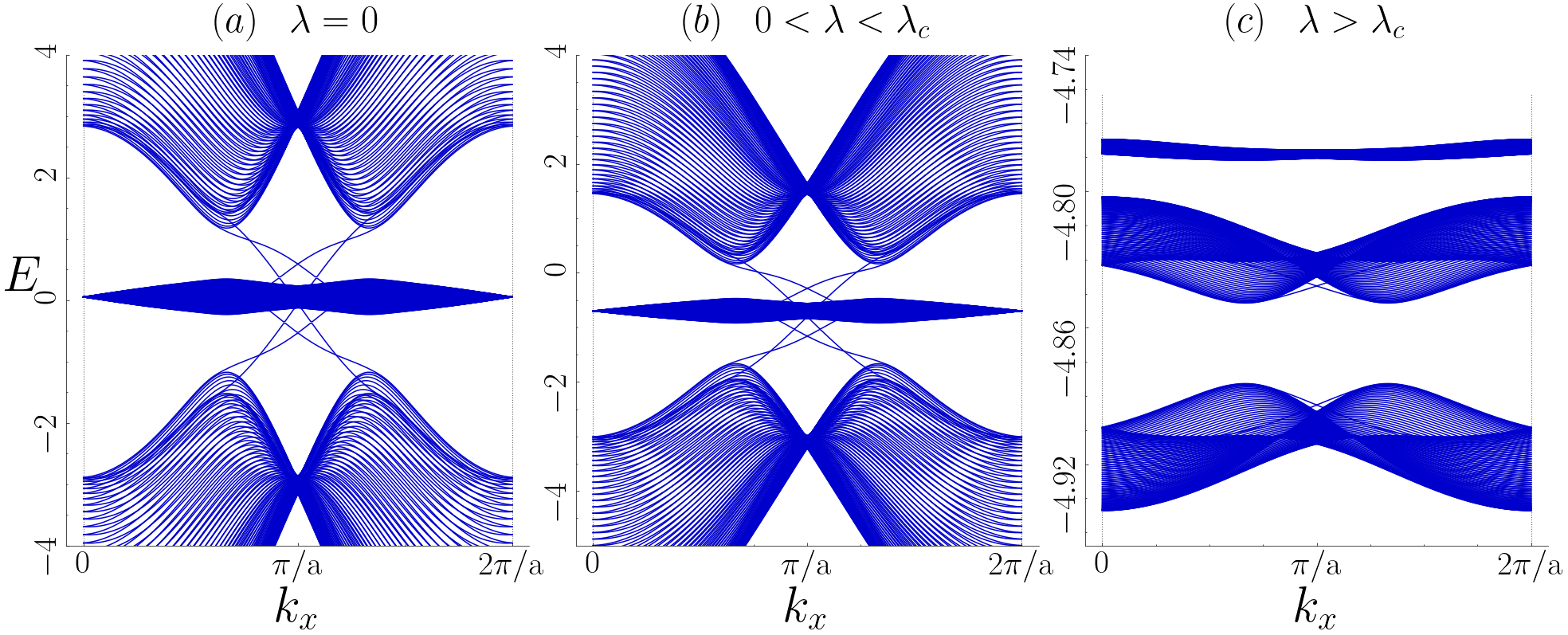}
\caption{Edge state spectra (in units of $t$) of a zigzag edged semi-infinite ribbon for $\alpha=0.2$ are shown as a function of dimensionless momenta, $k_x$ for $(a)$ $\lambda=0$, $(b)$ $\lambda=0.5$ ($0<\lambda<\lambda_{c}$), and $(c)$ $\lambda=2.2$ ($\lambda>\lambda_{c}$). Other parameters are the same as those in Fig.\,\ref{fig:bulk_p2}. The values of $\lambda_{c}$ are mentioned in Table \ref{tab:table1}.}
\label{fig:edgep2}
\end{figure} 
\begin{figure}
\includegraphics[width=\columnwidth]{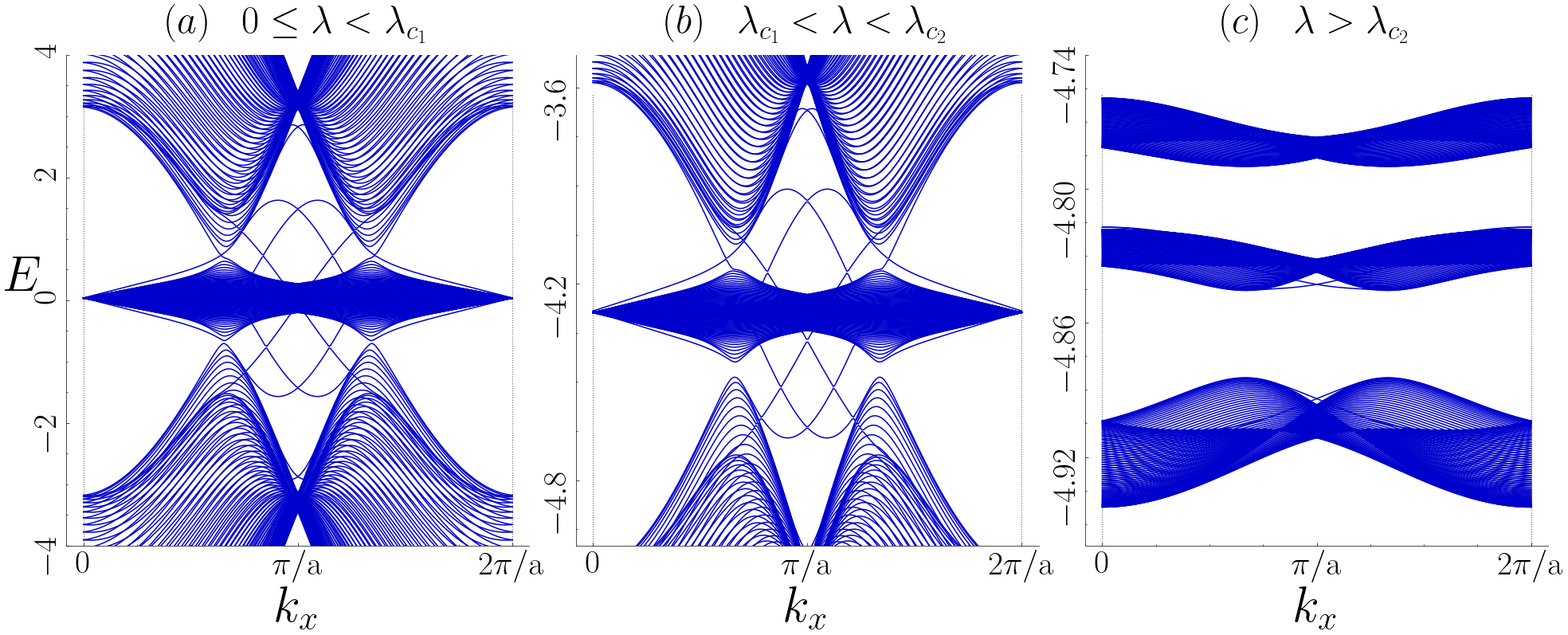}
\caption{Edge state spectra (in units of $t$) of a zigzag edged semi-infinite ribbon for $\alpha=0.52$ are shown as a function of dimensionless momenta, $k_x$ for $(a)$ $\lambda=0$ ($0\leq\lambda<\lambda_{c_1}$), $(b)$ $\lambda=1.2$ ($\lambda_{c_1}<\lambda<\lambda_{c_2}$), and $(c)$ $\lambda=2.2$ ($\lambda>\lambda_{c_2}$). Other parameters are the same as those in Fig.\,\ref{fig:bulk_p52}. The values of $\lambda_{c_1}$ and $\lambda_{c_2}$ are mentioned in Table \ref{tab:table1}.}
\label{fig:edgep52}
\end{figure} 
\begin{figure}
\includegraphics[width=\columnwidth]{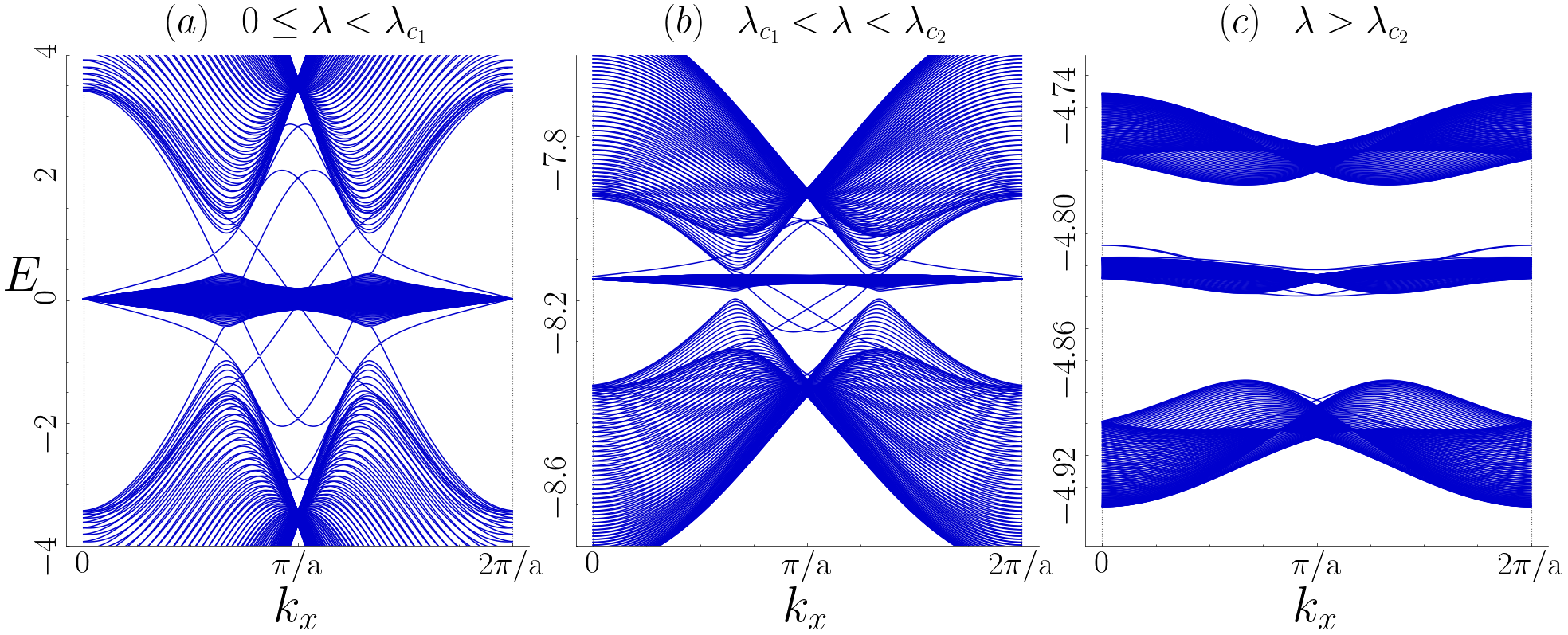}
\caption{Edge state spectra (in units of $t$) of a zigzag edged semi-infinite ribbon for $\alpha=0.7$ are shown as a function of dimensionless momenta, $k_x$ for $(a)$ $\lambda=0$ ($0\leq\lambda<\lambda_{c_1}$), $(b)$ $\lambda=1.65$ ($\lambda_{c_1}<\lambda<\lambda_{c_2}$), and $(c)$ $\lambda=2.2$ ($\lambda>\lambda_{c_2}$). Other parameters are the same as those in Fig.\,\ref{fig:bulk_p7}. The values of $\lambda_{c_1}$ and $\lambda_{c_2}$ are mentioned in Table \ref{tab:table1}.}
\label{fig:edgep7}
\end{figure}

This section is devoted to studying the effect of e-p coupling on the edge state characteristics of a semi-infinite $\alpha$-$T_{3}$ ribbon with zigzag edges~\cite{Alam2019}. In our case, the nanoribbon is infinite (protecting translational symmetry) along the $x$-direction, while finite along the $y$-direction, breaking the translational symmetry along $k_y$. To satisfy the condition of the width of a zigzag edge, written as $N=3q+1$ ($q$ is an integer), we have taken the width of the ribbon as $N=100$, which ensures that both the edges are composed of A and C sublattices only. Here, we investigate the appearance (vanishing) of the edge states in different regimes of $\lambda$ for the same three $\alpha$ cases as earlier, whose bulk spectra and $\mathbb{Z}_2$ invariant are inspected in Sec.~\ref{textbulk} and Sec.~\ref{textZ2}, respectively. The distinction between a topological and non-topological phase relies on the crossings of the edge modes between CB and VB through the FB. We look for a direct correspondence between the evolution of HWCCs ($\mathbb{Z}_2$ invariant) and the edge states. Conventionally, if the arbitrary line intersects the HWCCs an odd number of times in the half-BZ, the Fermi level also crosses an odd number of edge states per edge during this half-BZ evolution. 

We begin by referring to Fig.\,\ref{fig:edgep2}$(a)$, where we vividly notice the emergence of two pairs of the edge states traversing from the VB to CB through the FB at $\lambda=0$ for $\alpha=0.2$. These are helical QSH edge states of a $\mathbb{Z}_2$ topological insulator, which counterpropagate for two different spin ($\uparrow$ and $\downarrow$) bands with a velocity, $v=\partial E/\partial k$ (slope of the bands). However, we cannot distinctly designate them as the $\uparrow$-spin or the $\downarrow$-spin bands as the spins are mixed by the presence of the RSOC. These helical modes continue to persist in the $\lambda<\lambda_c$ regime, displayed in Fig.\,\ref{fig:edgep2}$(b)$. Therefore, for $0\leq\lambda<\lambda_c$, an energy lying between the FB and the VB (the \text{`so called'} Fermi level) intersects the edge states odd (one) number of times per edge for a particular spin component (giving rise to two in total) and in argument with an odd crossing in HWCCs (non-zero $\mathbb{Z}_2$ invariant) during the half-BZ evolution. We can reconcile this by the argument of TR \textit{polarization pumping}~\cite{FuI2006} as one unit of $\uparrow$-spin is pumped relative to the $\downarrow$-spin to the edge while HWCCs evolve through half-BZ giving rise to a finite conduction along the two edges. The existence of these helical edge modes in the $0\leq\lambda<\lambda_c$ regime directly corresponds to the gapped (topologically nontrivial) phase in the bulk (Figs.\,\ref{fig:bulk_p2}$(a)$ and $(b)$) for which the $\mathbb{Z}_2$ invariant is found to be (odd) non-zero (Fig.\,\ref{fig:Z2p2}$(a)$ and $(b)$) clarifying the system to be a $\mathbb{Z}_2$ topological insulator. As anticipated earlier, both by the bulk diagrams and the $\mathbb{Z}_2$ invariants, beyond $\lambda=\lambda_c$, these edge states disappear (Fig.\,\ref{fig:edgep2}$(b)$) completely (even cut in the HWCCs in Fig.\,\ref{fig:Z2p2}$(c)$), inferring the system to be a trivial insulator. 
It is evident by the bulk spectra (Fig.\,\ref{fig:bulk_p52}$(b)$) that there is an $SM$ point at $\alpha=0.52$ for the $0\leq\lambda<\lambda_{c_1}$ (including $\lambda=0$ shown in Fig.\,\ref{fig:bulk_p52}$(a)$) regime which also reflects in the edge spectrum, presented in Fig.\,\ref{fig:edgep52}$(a)$. The concept of edge states is, therefore, not so important for this regime of $\lambda$ as the bulk bands are gapless. However, as we tune $\lambda$ further, the two pairs of the helical edge states emerge which traverse from the FB to the VB (Fig.\,\ref{fig:edgep52}$(b)$) and thus yield the signature of a topological QSH insulator (also seen through an odd number of cut in HWCCs shown in Fig.\,\ref{fig:Z2p2}$(a)$). Nevertheless, the vanishing of edge states in the $\lambda>\lambda_{c_2}$ (Fig.\,\ref{fig:edgep52}$(c)$) confirms this phase to be a trivial insulator.   
For a specific case of higher $\alpha$ value, say $\alpha=0.7$, it is shown in Fig.\,\ref{fig:edgep7}$(a)$ that although edge states exist in the $0\leq\lambda<\lambda_{c_1}$ regime (where the bulk is gapped represented in Fig.\,\ref{fig:bulk_p7}$(a)$ and $(b)$), however a total of four pairs appear, which essentially implies two pairs of \text{`copropagating'} edge modes along a particular edge.  Each of the two pairs may be denoted by the two spin orientations, giving rise to a possibility of backscattering between the same spin states, which destroys the robustness of a topological phase. Consequently, the \text{`Fermi level'} crosses them eight times which manifests a non-topological phase with zero conduction along edges (also confirmed by an even cut in the HWCCs (Fig.\,\ref{fig:Z2p7}$(a)$)). Interestingly, the gapped spectrum in the $\lambda_{c_1}<\lambda<\lambda_{c_2}$ regime comprises of two pairs of helical edge states (Fig.\,\ref{fig:edgep7}$(b)$) which carries a trait of a $\mathbb{Z}_2$ QSH insulator characterized by an odd number of crossing in the HWCCs (Fig.\,\ref{fig:Z2p7}$(b)$). As these edge states are susceptible to the e-p coupling, we further enhance $\lambda$ and find that they entirely vanish (agreeing with an even cut in the HWCCs (Fig.\,\ref{fig:Z2p7}$(b)$)) for $\lambda>\lambda_{c_2}$ (Fig.\,\ref{fig:edgep7}$(c)$) where the system behaves as a trivial insulator. We want to assert that the results for lower ($0<\alpha\leq 0.5$) and higher ($0.6\leq\alpha<1.0$) values of $\alpha$ are identical to those for $\alpha=0.2$ and $\alpha=0.7$, respectively. However, we do not show all of them here for brevity. Hence, the evolution of the HWCCs giving rise to a $\mathbb{Z}_2$-odd (even) invariant consistent with the emergence (vanishing) of the edge states corroborates that a pseudospin-$1$ fermionic system, such as, an $\alpha$-$T_3$ lattice goes through a topological-trivial and a trivial-topological-trivial phase transition for the $0<\alpha\leq 0.5$ and $0.6\leq\alpha<1.0$ regimes, respectively, with an intervening $SM$-topological-trivial transition occurring at $\alpha=0.52$. 

\subsection{The phase boundary}\label{textphase}
\begin{figure}
\centering
\includegraphics[width=0.85\linewidth]{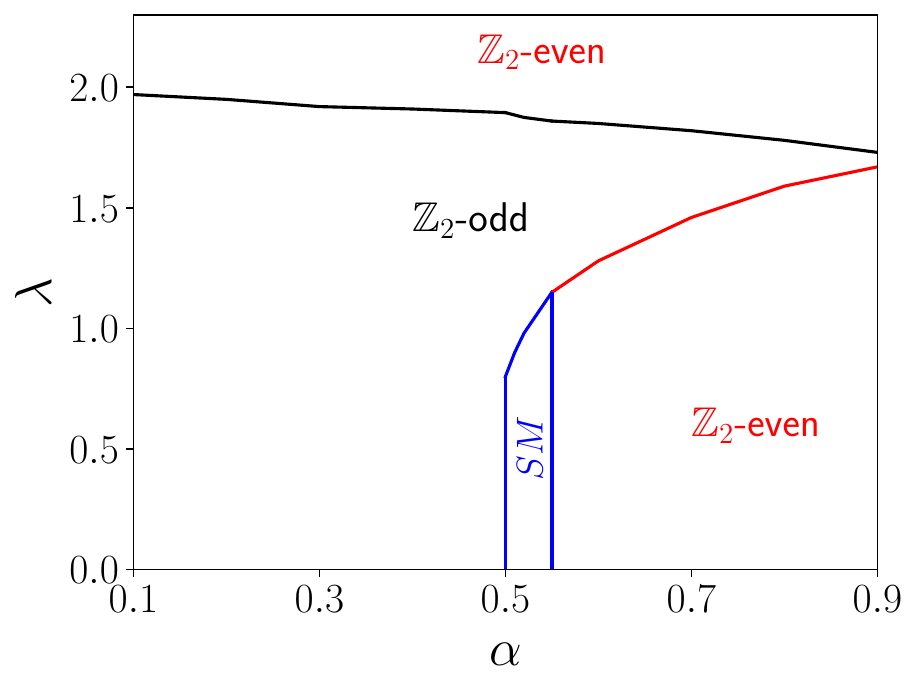}
\caption{The phase boundary of $\mathbb{Z}_2$ invariant in the $\alpha$-$\lambda$ plane. Various phases with even (trivial) and odd (topological) $\mathbb{Z}_2$ values and an $SM$ phase are shown.}
\label{fig:phase}
\end{figure}

It may be a good idea to depict all the distinct phases, namely, the topological, $SM$, and the trivial phases in the form of a phase diagram which encapsulates our findings in different regimes of e-p coupling ($\lambda$) for all values of $\alpha$ in the range $[0:1]$. To do the same, we show the $\mathbb{Z}_2$ invariant in the $\alpha$-$\lambda$ plane, presented in Fig.\,\ref{fig:phase}. The phase boundaries are basically the coordinates of the critical transition points $\lambda_c$ which demarcate the topological characteristics of the phases for different combinations of $\alpha$ and $\lambda$. In the figure, the black boundary denotes that below a critical $\lambda$ (see Table \ref{tab:table1} for $\lambda_c$ values), the system with lower values of $\alpha$ ($0<\alpha\leq 0.5$) possesses a $\mathbb{Z}_2$-odd topological phase which ceases to exist above values of $\lambda_c$ where the system acquires a $\mathbb{Z}_2$-even trivial phase. Beyond $\alpha=0.5$, there exits a tiny region (for $0.51<\alpha<0.54$) where the system behaves like $SM$ accompanied by a gapless dispersion ($\mathbb{Z}_2$ being ill-defined) below a certain $\lambda_{c_1}$ marked by the blue line. Above $\lambda_{c_1}$, it enters into a $\mathbb{Z}_2$-odd topological phase (the region between the blue and black boundaries) which remains unaltered till $\lambda$ reaches another critical point $\lambda_{c_2}$ beyond which the system becomes an ordinary ($\mathbb{Z}_2$-even) insulator. As encountered in the preceding subsections, the higher $\alpha$ cases demonstrate two bulk closing transitions at $\lambda_{c_1}$ and $\lambda_{c_2}$, which are shown by red and black boundaries, respectively in Fig.\,\ref{fig:phase} as a function of $\alpha$.
The regions enclosed by these boundaries of the phase diagram reveal that our system, for higher $\alpha$ values ($0.54<\alpha\leq 0.9$) acts as a trivial insulator below $\lambda=\lambda_{c_1}$ (region beneath the red curve) which is designated by the $\mathbb{Z}_2$-even region, while it shows the fingerprints of a topological insulator in the $\lambda_{c_1}<\lambda<\lambda_{c_2}$ regime, labelled by the $\mathbb{Z}_2$-odd region. It is obvious that for $\lambda>\lambda_{c_2}$ the system turns out to be a trivial ($\mathbb{Z}_2$-even) insulator for any value of $\alpha$ in the range $0.54<\alpha\leq 0.9$. It is also evident from Fig.\,\ref{fig:phase} that as we increase $\alpha$, the $\lambda_c$ values (black line) decrease in the range $0<\alpha\leq 0.5$, and also the difference between $\lambda_{c_1}$ and $\lambda_{c_2}$ (width of the region bounded by the red and black curves) diminishes for $0.5<\alpha\leq 0.9$.
  
So far, we have discussed the first order topology which manifests interesting transitions induced by the polaronic effects. It will be shown in the following that richer physics awaits if the TRS of the QSH insulator is broken, and the resultant phase shows intriguing phase transitions, including an HOTI phase. Specifically, a fully aperiodic system reveals accumulation of corner modes, and to the best of our knowledge, such SOTI has never been reported in an $\alpha$-$T_3$ lattice.  
\section{Emergent second order topological phase and relative transitions}\label{textSOTI}
While HOTI has gained considerable interest in the field of topology as a novel phase of matter, it is still not clear how the edge states of a TI can be efficiently gapped out to give rise to higher order topological states.
Several mechanisms, one of which involves the inclusion of orbital currents that break the TRS distinctly in the $x$ and $y$-directions, in the celebrated Bernevig-Hughes-Zhang (BHZ) model, consequently giving rise to corner states, has been elaborately discussed and explored in several studies \cite{Schindler2018}.
Furthermore, the Benalcazar-Bernevig-Hughes (BBH) model, which studies a 2D bipartite square lattice system with an additional $\pi$-flux per plaquette (owing to the presence of a magnetic field that penetrates the system in the $z$-direction), also describes a well defined HOTI phase characterized by a robust quadrupole moment \cite{Benalcazar2017}.
It was also observed in Ref.~\cite{Lahiri2024} and Ref.~\cite{LahiriII2024} that inculcating a uniaxial strain in the Haldane and the Kane Mele model leads to the destruction of first order topology while giving rise to resilient second order corner states beyond a critical value of the strain is surpassed.
In this regard, Ren~\etal~\cite{Ren2020} explored another possibility of arriving at an HOTI phase, by studying the Kane Mele model under the action of an in-plane magnetic field that breaks the TRS and hence disturbs the $\mathbb{Z}_2$ topology.
It was observed that the magnetic field gaps out the edge states in a $x$-periodic (zigzag) ribbon, thus giving rise to corner states on a suitable rhombic geometry characterized by a mirror-graded winding number.

Along the lines as above, we employ a similar in-plane magnetic field to trigger the exploration of an HOTI phase, or more precisely, an SOTI phase in the $\alpha$-$T_3$ lattice which resemblances a pseudospin-$1$ fermionic system hosting FBs and subsequently observe how this emergent higher order topology responds to the e-p coupling.
The Hamiltonian under the action of the in-plane magnetic field directed along the $x$-direction can be written as
\begin{equation}
    \mathcal{H_\text{SOTI}} = \tilde{\mathcal{H}}_{\text{eff}} + \mathcal{H_{\text{B}}},
\end{equation}
where,
\begin{equation}
\mathcal{H_\text{B}}=B_x\sigma_x\otimes I_3.
\end{equation}
Here, $B_x$ corresponds to the strength of the magnetic field.
\begin{figure}
\centering
\includegraphics[width=\columnwidth]{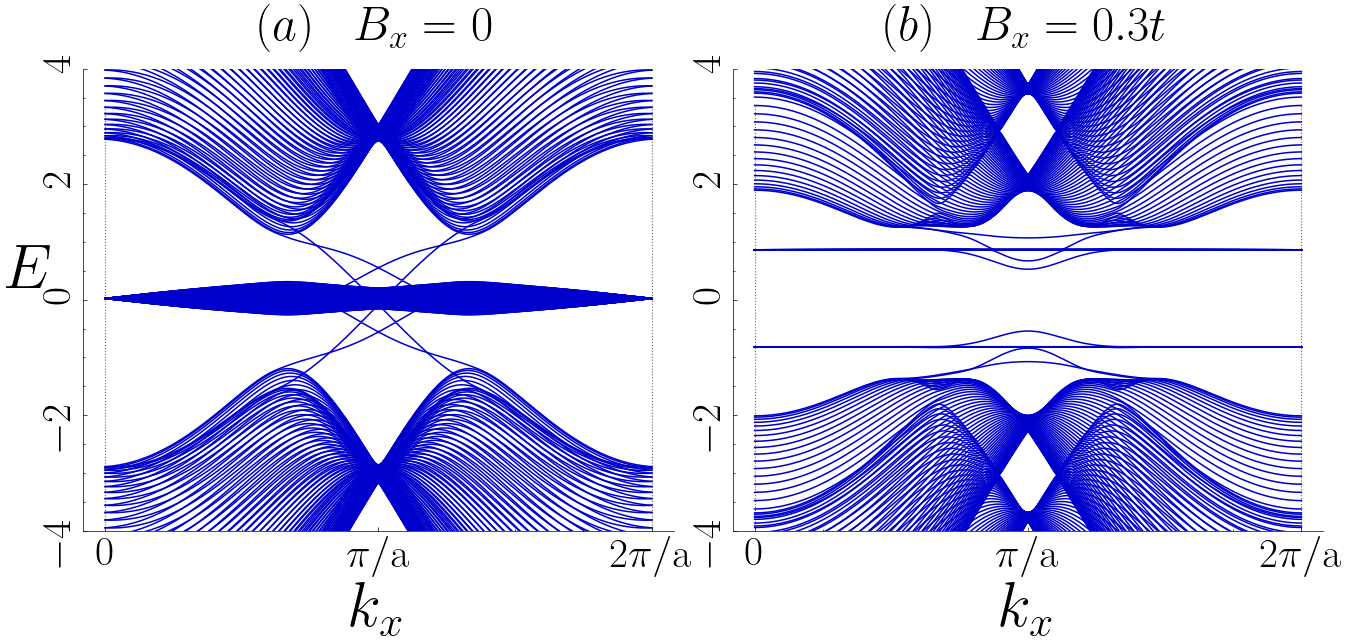}
\caption{The edge spectra (in units of $t$) of the nanoribbon configured with periodicity along the $x$-direction (zigzag edges) is shown for $\alpha=0.1$ and $\lambda=0.1$. In $(a)$ for $B_x=0$, the edge state are protected by TRS (QSH states), and in $(b)$ for $B_x=0.3t$, the edge modes gap out on the introduction of the magnetic field, indicating an apparent destruction of the first order topology.}
\label{HOTI1}
\end{figure}
$\sigma_x$ and $I_3$ correspond to the spin and sublattice degrees of freedom, respectively, where $\sigma_x$ represents the $x$-component of the Pauli matrices and $I_3$ refers to a $3\times 3$ identity matrix.
$\tilde{\mathcal{H}}_{\text{eff}}$ has been obtained in Eq.~(\ref{Ham: eff model}).
\begin{figure}
\centering
\includegraphics[width=\columnwidth]{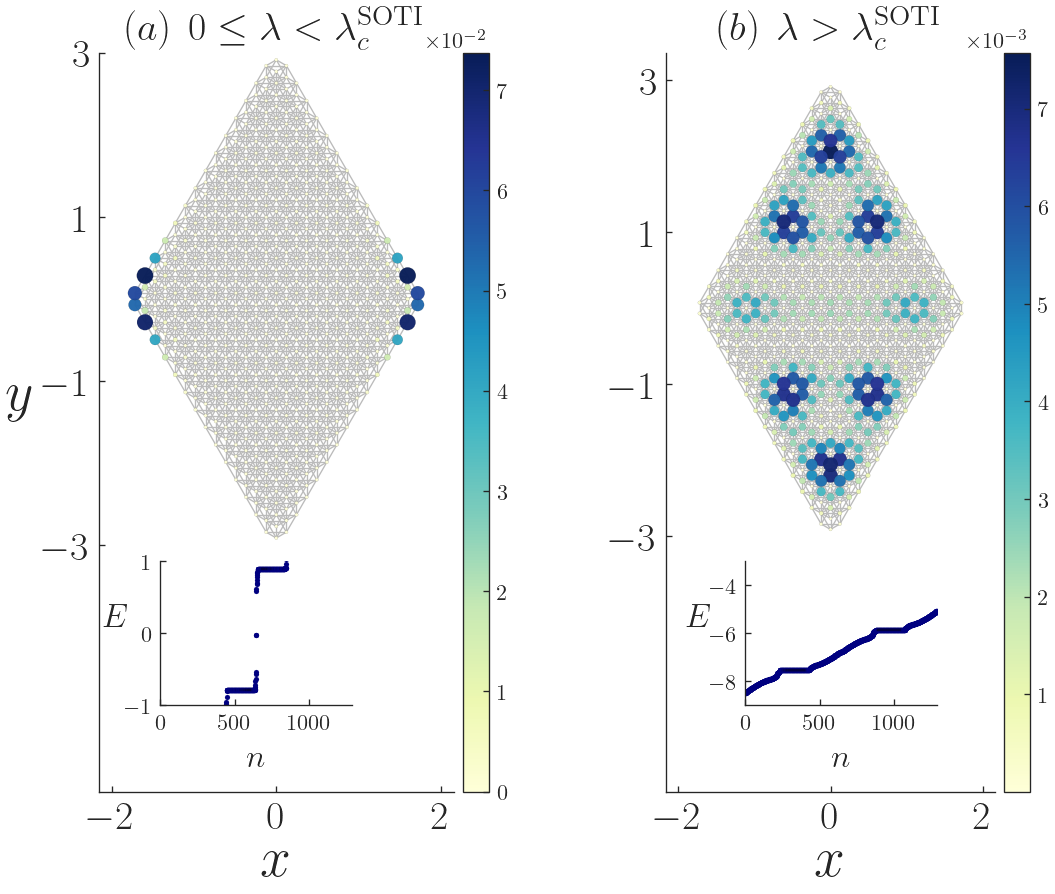}
\caption{$(a)$ The probability distribution of the in-gap states obtained due to the introduction of an in-plane magnetic field ($B_x$) in the $\alpha$-$T_3$ QSH Hamiltonian for $\alpha=0.1$ shows the emergence of a second order topology at $\lambda=0$. The magnetic field is fixed at $B_x=0.3t$. The second order states are obtained at the corners where two zigzag edges intersect. The inset shows the energy eigenvalues where the in-gap states are clearly visible. Here $n$ represents the state index. $(b)$ The confinement of the second order states is disturbed as soon as the e-p coupling surpasses a critical value, as represented in Fig.~\ref{HOTI2}. The value of $\lambda $ is taken to be $\lambda=1.5$ in this case.}
\label{HOTI4}
\end{figure}
Without loss of generality, the magnetic field has been taken to point in the $x$-direction for our study (a $y$-directed field causes no difference~\cite{Ren2020}).
It is observed that on the introduction of the magnetic field, the first order edge states for a zigzag nanoribbon of the $\alpha$-$T_3$ lattice (with $k_x$ as the periodic parameter) instantly gap out (Fig.~\ref{HOTI1}).
This indicates an apparent destruction of the first order topological phase.
\begin{figure}
\centering
\includegraphics[width=\columnwidth]{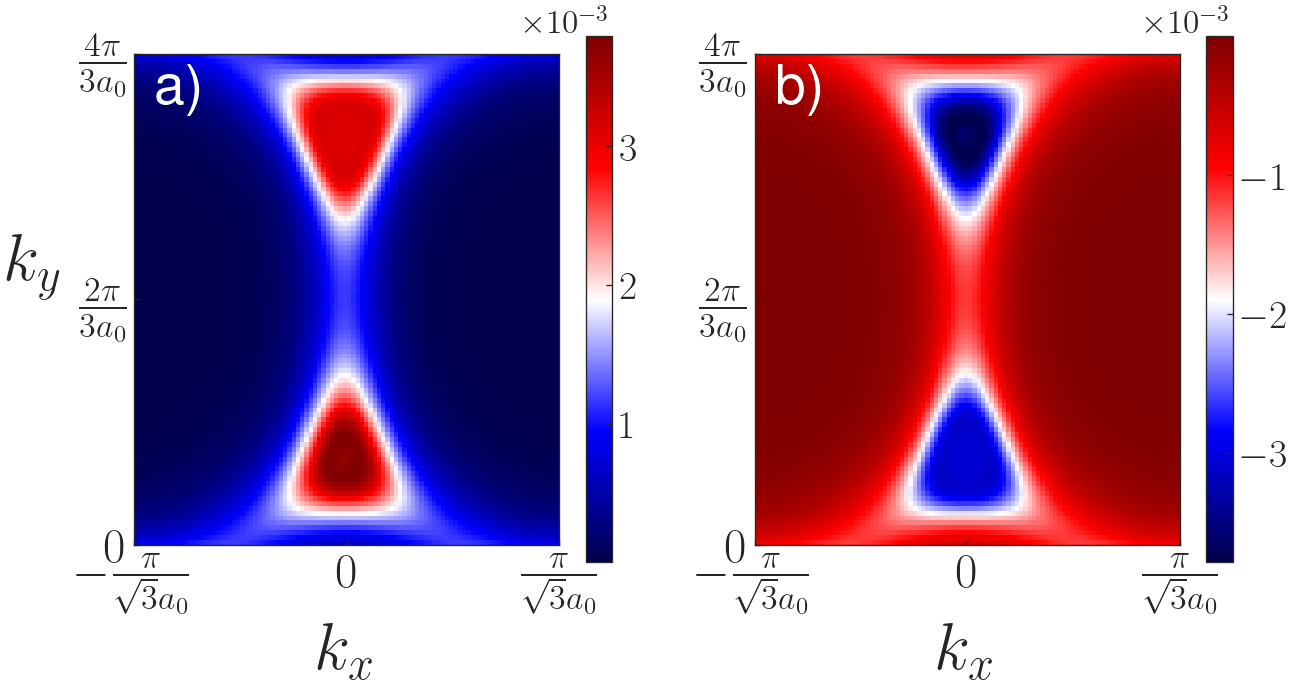}
\caption{The spin Berry curvature pertaining to the $(a)$ projected $\uparrow$-spin and $(b)$ $\downarrow$-spin states is calculated using the eigenvectors $\eta_1$ and $\eta_6$ of the projected spin operator $S$ for $\alpha=0.1$ and $\lambda=0.1$ ($0\leq\lambda<\lambda_c^{\mathrm{SOTI}}$). The corresponding Chern numbers are calculated to be $+1$ and $-1$, respectively. The in-plane magnetic field is fixed at $B_x=0.3t$, while the RSOC has been kept at zero.}
\label{HOTI3}
\end{figure}
\begin{figure}
\centering
\includegraphics[width=\columnwidth]{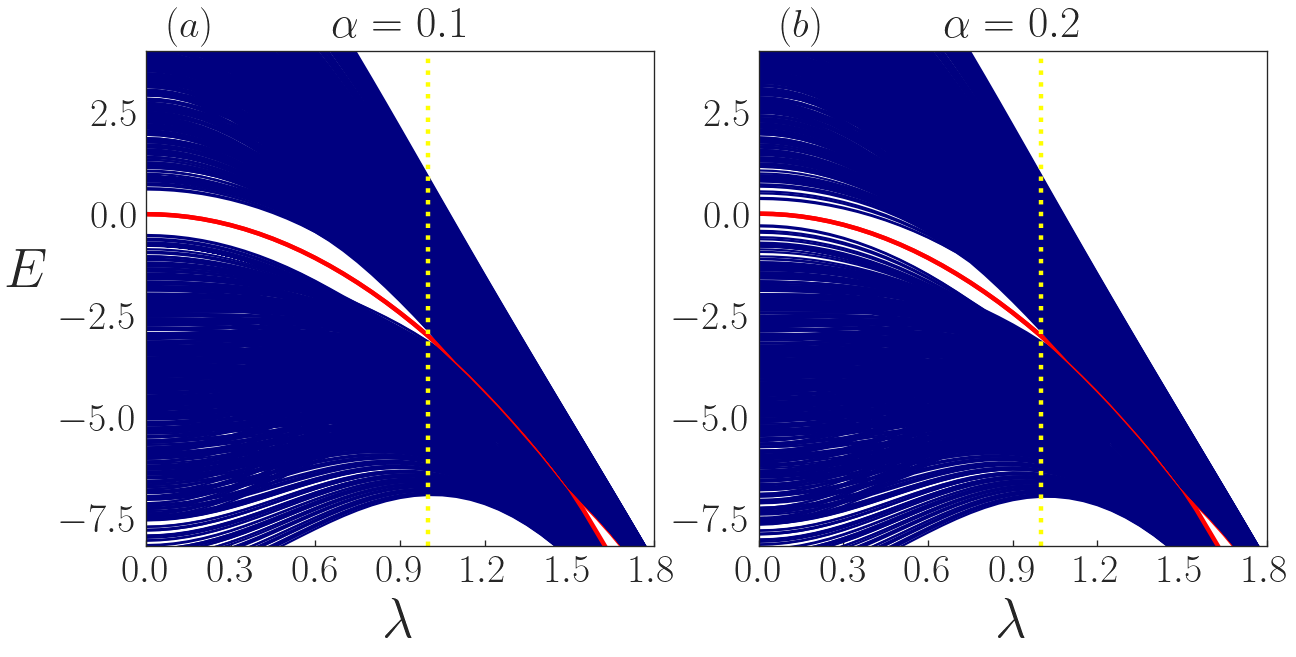}
\caption{The real space energy spectra (in units of $t$) of the $\alpha$-$T_3$ QSH Hamiltonian projected on a rhombic geometry has been plotted as a function of the e-p coupling $\lambda$ for $(a)$ $\alpha=0.1$ and $(b)$ $\alpha=0.2$. The dotted line indicates the value of $\lambda$ at which the in-gap states merge into the bulk of the system. Here $B_x$ is kept fixed at $B_x=0.3t$.}
\label{HOTI2}
\end{figure}
However, as discussed in Ref.~\cite{Ren2020} and Ref.~\cite{halder2024}, the in-plane field is not capable of destroying the bulk topology and transforming the system into a trivial insulator.
In the Kane-Mele model (or BHZ), it is only capable of gapping out the conducting states in a zigzag ($x$-periodic) edge while maintaining them intact in an armchair ($y$-periodic) edge.  

With this in mind, the $\alpha$-$T_3$ Hamiltonian in the presence of an in-plane magnetic field is now projected on a rhombic supercell with zigzag edges to observe the fate of the boundary states on a completely open geometry. In our study, the strength of the magnetic field $B_x$ has been taken to be $0.3t$ to show the SOTI transition for $\alpha=0.1$ and $\alpha=0.2$ and the RSOC is turned off as the latter has no significant role in the context of HOTI. However, we keep the strength of KM term as $t_{so}=0.1t$ throughout this section.  We assert that the only significance of the magnetic field is to break the TRS (for more details see below). Further, it yields a boundary dependent topological features, where the armchair ribbon continues to support robust edge modes~\cite{Ren2020}.

The emergence of the in-gap states is clearly observed with the probability density confined at those corners where two zigzag edges intersect (Fig.~\ref{HOTI4}$(a)$), signalling an occurrence of a second order topology. Moreover, Fig.~\ref{HOTI4}$(b)$ demonstrates that there exists a critical $\lambda$, namely $\lambda_c^{\mathrm{SOTI}}$ (corresponding in-gap states are explicitly shown in Fig.~\ref{HOTI2} via the red line), beyond which the confinement of the corner states is disrupted. 
The occurrence of these higher order corner states arises solely from the fact that the magnetic field is only successful in gapping out the edge states of the zigzag nanoribbon, while maintaining the bulk topology intact.

To verify the above claim, we evaluate the projected spin Chern number for a point in the parameter space where second order topology is endorsed by the system \cite{Yang2011, Prodan2009}.
This is done since the $\mathbb{Z}_2$ invariant fails to characterize topology in presence of a magnetic field.
It is to be noted that a small value of $B_x$ has been chosen throughout our study to ensure that the split FB is well separated in energy from the VB, thus obstructing hybridization and ensuring that the projected spin Chern number stays well defined. 
Given that the spin operator for the $\alpha$-$T_3$ model is represented as $\hat O=\sigma_z\otimes I_3$, the projected spin operator $S$ can be defined as \cite{Saha2021}
\begin{equation}
    S(\mathbf{k})=P(\mathbf{{k}})\hat OP(\mathbf{{k}}),
\end{equation}
where $P(\mathbf{k})$ corresponds to the projection operator on the VB and is given as
\begin{equation}
P(\mathbf{{k}})=|u_1(\mathbf{k})\rangle\langle u_1(\mathbf{k})|+|u_2(\mathbf{k})\rangle\langle u_2(\mathbf{k})|,
\end{equation}
where $|u_1(\mathbf{k})\rangle$ and $|u_2(\mathbf{k})\rangle$ correspond to the valence eigenstates of the Hamiltonian $\mathcal{H_\text{SOTI}}$.
The diagonalization of $S$ gives six eigenvalues $E_i$ ($i\in[1, 6]$). Further, $|E_1|=|E_6|=1$ corresponding to the eigenvectors $\eta_1$ and $\eta_6$, the rest of them being zero.
The spin Chern number $C_{\sigma}(\sigma=\uparrow,\downarrow)$ is now evaluated using the eigenvectors $\eta_1$ and $\eta_6$ of the projected spin operator $S$, employing the Fukui's method \cite{Fukui2005}.
Furthermore, the spin Berry curvature $\Omega_{\uparrow/\downarrow}$ is shown in Fig.~\ref{HOTI3}, where \cite{Fukui2005, Avron1988}
\begin{align}
    \begin{split}
       \Omega_{\sigma}(\mathbf{k}) &=i\biggl{[} \biggl <\frac{\partial \eta(\mathbf k)}{\partial k_x}\biggl|\frac{\partial \eta(\mathbf k)}{\partial k_y}\biggl> - \biggl <\frac{\partial \eta(\mathbf k)}{\partial k_y}\biggl|\frac{\partial \eta(\mathbf k)}{\partial k_x}\biggl>\biggl ].
    \end{split}
\end{align}
Here, $\eta\in \{\eta_1, \eta_6\}$ corresponding to the projected $\uparrow$-spin and $\downarrow$-spin states respectively.
It is to be noted that integration of the Berry curvature over the entire 2D Brillouin zone represents an alternative method of arriving at the spin Chern number.
Importantly, we observe that in spite of the presence of a magnetic field in the system, the spin Chern number for the projected up (down) spin states within the VB remains equal to 1 (-1), implying the in-plane field only gaps out the edge states of a zigzag nanoribbon without disturbing the bulk topology. Therefore, 
this non-trivial bulk topology is ultimately responsible for the emergence of the SOTI phase associated with the corner modes under a completely open boundary condition which are characterized in Fig.~\ref{HOTI3}.
It is important to note that the intersection between the two zigzag edges mimics a portion of an armchair boundary, thus hosting boundary states that resembles a higher order phase.
This justifies the correspondence between the projected spin Chern number and the corner modes in the rhombic supercell. Nevertheless, the prescription of the spin Chern number is not valid for $\lambda>\lambda_c^{\mathrm{SOTI}}$ as the bands merge into the bulk in this regime of $\lambda$ which we discuss below.

As mentioned already, the SOTI phase in our system is generated entirely by the in-plane magnetic field ($B_x$), which implies that the e-p coupling ($\lambda$) does not take part in generating these higher order states. This is clearly displayed in Fig.~\ref{HOTI1}$(a)$ which consolidates the validation of the corner states also for $\lambda=0$. Now, we want to investigate the role of the e-p coupling on the stability of the SOTI phase. Our primary motivation being the study of the variation of the SOTI phase in an $\alpha$-$T_3$ lattice with respect to the e-p coupling, we plot the real space bandstructure (obtained on projecting the Hamiltonian on the rhombic supercell as represented in Fig.~\ref{HOTI4}) as a function of $\lambda$ (Fig.~\ref{HOTI2}).
It is observed that the in-gap states survive till a critical value of $\lambda$ ($=\lambda_c^{\mathrm{SOTI}}$) before merging into the bulk of the system.
The probability density of the states that were earlier confined to the corners of the supercell are now distributed all over the lattice implying onset of a trivial behaviour, when $\lambda$ exceeds $\lambda_c^{\mathrm{SOTI}}$ (Fig.~\ref{HOTI4} (b)).
The applied magnetic field $B_x$ (=$0.3t$) causes the SOTI-trivial transition around $\lambda_c^{\mathrm{SOTI}}=1.1$ for $\alpha=0.1$ and $\alpha=0.2$.
For lower values of $\lambda$ (within the regime $0\leq\lambda<\lambda_c^{\mathrm{SOTI}}$), the energy of the in-gap states remains close to $E=0$, while it shifts away for higher values of $\lambda$.
The degeneracy of the states is however maintained intact.
Furthermore, the dependence of the value of $\lambda_c^{\mathrm{SOTI}}$ on the B-C coupling parameter $\alpha$ is weak, as shown in Fig.~\ref{HOTI2}.

\begin{figure}
\centering
\includegraphics[width=\columnwidth]{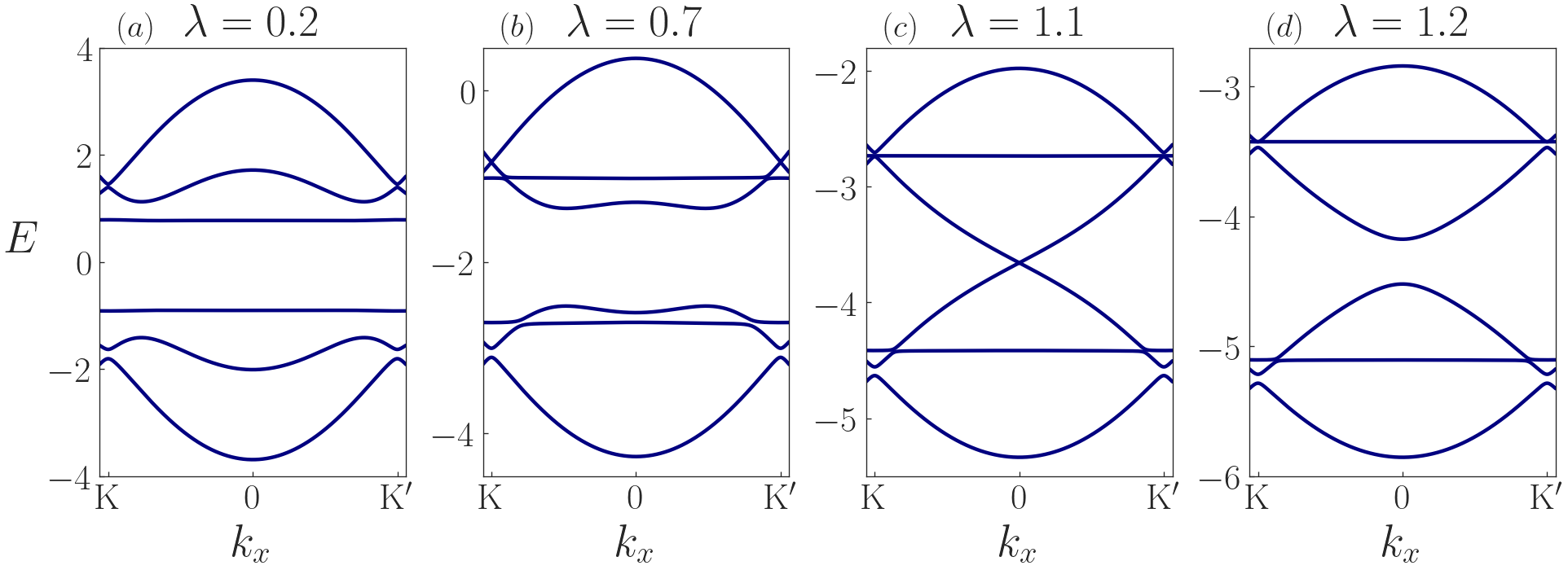}
\caption{The bulk states (in units of $t$) for the $\alpha$-$T_3$ QSH Hamiltonian is plotted for $\alpha=0.1$ as a function of $k_x$ while keeping $k_y$ fixed at $\frac{2\pi}{3a_0}$ ($a_0$ being the NN distance) for $(a)$ $\lambda=0.2$, $(b)$ $\lambda=0.7$, $(c)$ $\lambda=1.1$ and $(d)$ $\lambda=1.2$. It is observed that the system undergoes a bulk gap closing transition around $\lambda=\lambda_c^{\mathrm{SOTI}}=1.1$, which closely corresponds to the point where the in-gap states in the real space bandstructure merge into the bulk destroying the SOTI phase. We fix $B_x=0.3t$.}
\label{HOTI5}
\end{figure}

\begin{figure}
\includegraphics[width=\columnwidth]{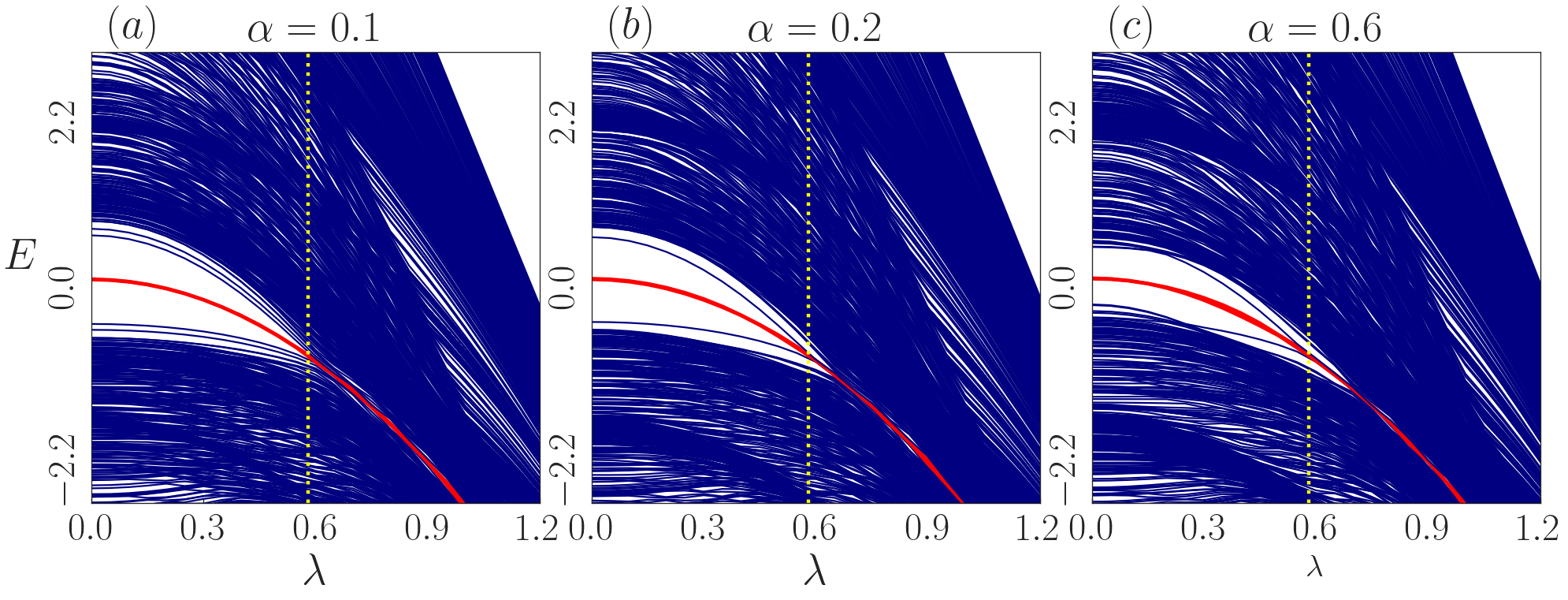}
\caption{The real space energy spectra (in units of $t$) of the $\alpha$-$T_3$ QSH system have been plotted as a function of the e-p coupling $\lambda$ for $(a)$ $\alpha=0.2$, $(b)$ $\alpha=0.4$ and $(c)$ $\alpha=0.6$. The magnetic field strength, $B_x$, has been taken to be equal to $0.7t$ such that the SOTI phase persists for higher values of $\alpha$.}
\label{HOTIA1}
\end{figure}

We now resort to study the bulk bandstructure which also shows intricate features.
The introduction of a magnetic field splits the bulk energy bands, which are otherwise degenerate in the absence of $\alpha$ and $\lambda$.
A strong magnetic field has the effect of pushing the two FBs away from each other which is distinct from the effect that either $\lambda$ or $\alpha$ has on the FBs.
In the presence of the magnetic field, a gap closing transition is clearly seen to occur as a function of $\lambda$ in the bulk bandstructure close to $\lambda_c^{\mathrm{SOTI}}$.
This clearly hints at a topological phase transition that destroys the SOTI phase (Fig.~\ref{HOTI5}).
The $x$-periodic edge spectrum in this regime ($\lambda>\lambda_c^{\mathrm{SOTI}}$) hints at a semi-metallic phase with no indication of the presence of distinct edge states.
Therefore, a very clear correspondence exists between the bulk, nanoribbon ($x$-periodic) and the open boundary spectra at $\lambda=\lambda_c^{\mathrm{SOTI}}$, where all the three spectra undergo a distinct transition.
Thus, e-p coupling serves as an appropriate tool for inducing higher order TPT in an $\alpha$-$T_3$ lattice. We should note that the original e-p coupling strength ($g$) at which these SOTI transitions take place can be obtained through the relation $\lambda=g^2\omega_0/t$, as mentioned in Sec.~\ref{textFOTI}.\\
\textit{Interplay of $B_x$ and $\alpha$:} Finally, we dwell on the role of $\alpha$ in generating the SOTI phases and the relevant topological transitions. The objective is to understand the distinct features of a pseudospin-$1$ fermionic $\alpha$-$T_3$ lattice (hosting FBs for non-zero $\alpha$) in the context of an SOTI. 
It is observed that the emergence of the SOTI phase is precluded unless the strength of the in-plane field ($B_x$) surpasses the value of $\alpha$.
This is shown in Fig.~\ref{HOTIA1} where the value of the magnetic field has been fixed at $0.7t$ (such a high value of $B_x$ has been deliberately chosen to show the existence of an SOTI phase for higher $\alpha$).
Clearly, the in-gap states persist for larger values of $\alpha$.
This implies that to obtain distinct in-gap corner states in the regime of higher $\alpha$, the magnetic field has to be concurrently enhanced. 
Therefore, for an $\alpha$-$T_3$ lattice, the choice for the value of the magnetic field to exhibit the SOTI phase is not unique.
Instead, it is highly correlated with the parameter $\alpha$. We should also mention that the value of the critical $\lambda$ at which the in-gap states merge into the bulk significantly decreases when $B_x$ is large (see Fig.~\ref{HOTIA1}).
\begin{figure}
\includegraphics[width=\columnwidth]{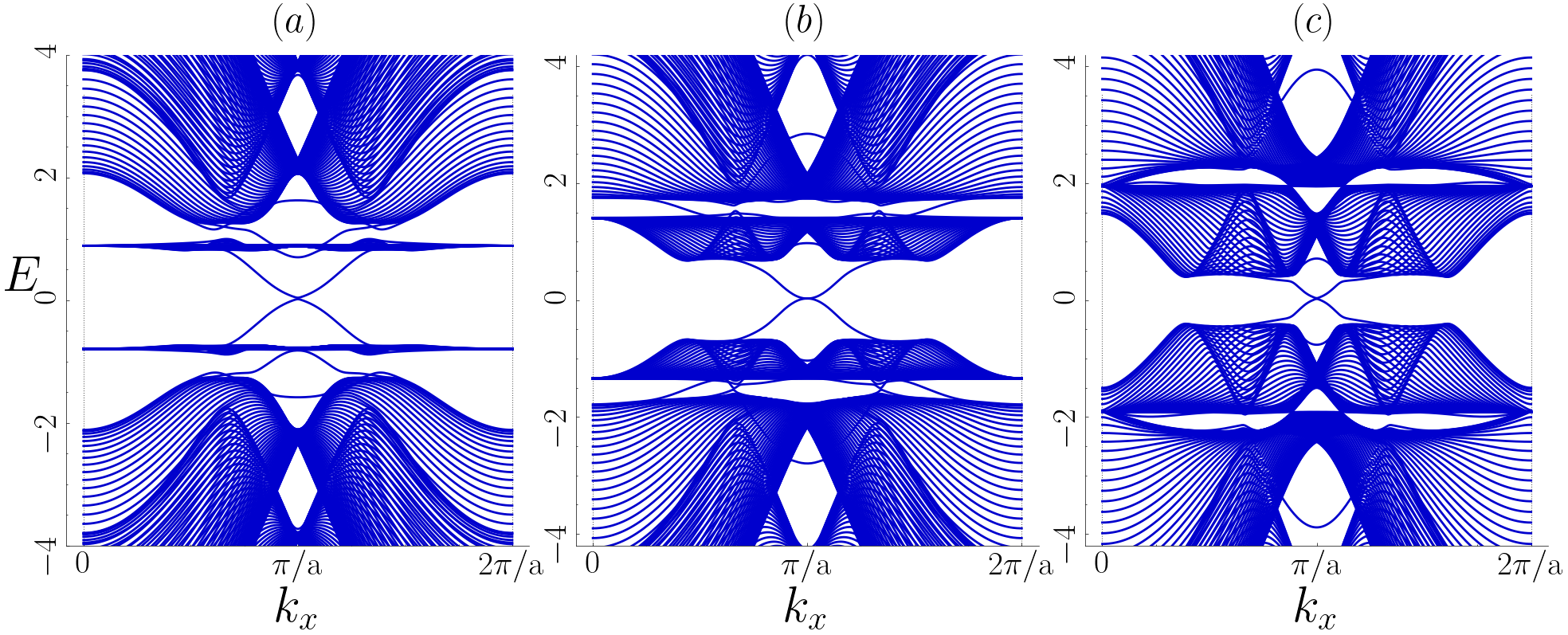}
\caption{The bandstructure (in units of $t$) of the zigzag nanoribbon for $(a)$ $\alpha=B_x=0.3$, $(b)$ $\alpha=B_x=0.5$ and $(c)$ $\alpha=B_x=0.7$. The value of $\lambda$ has been kept fixed at zero. It is evident that the gap which is opened in the spectrum by the application of $B_x$ closes as soon as $\alpha$ matches $B_x$.}
\label{HOTIA2}
\end{figure}
Having ascertained the SOTI transitions by e-p coupling, we now estimate the range of $\alpha$ to host the SOTI phases, for a given in-plane magnetic field.
The interplay between $B_x$ and $\alpha$ is understood more succinctly by studying the bandstructure of the zigzag nanoribbon, explicitly considering the points where $\alpha\approx B_x$, which is shown in Fig.~\ref{HOTIA2}.
The e-p coupling is fixed at zero for this study so as to extricate the effects of only $B_x$ and $\alpha$ on inducing the SOTI phase.
We observe that the spectrum (of the zigzag ribbon), which was erstwhile gapped out due to the introduction of the magnetic field, undergoes a gap closing when $\alpha\approx B_x$.
This phenomenon suggests that we can tune $B_x$ up to $B_x\approx\alpha$ to exhibit the formation of an SOTI phase for any $\alpha$ in range $[0:1]$ (validating the interpolation between graphene ($\alpha=0$) and dice ($\alpha=1$) lattices).
Hence, the inability to host the SOTI phase as $\alpha$ exceeds $B_x$ can be explained by this gap closing, thus emphasizing the role of $\alpha$ in maneuvering the second order topology in an $\alpha$-$T_3$ lattice.
\section{Conclusion}\label{Sec:summary}
To conclude, we study the effects of e-p interaction on inducing a FOTPT and SOTPT in a pseudospin-$1$ fermionic system on an $\alpha$-$T_3$ lattice. We formulate our model Hamiltonian in the spirit of the Kane-Mele Model modified by the Holstein term accounting for the e-p interaction and also include the effects of RSOC. The Lang-Firsov approach (to capture polaron physics in the anti-adiabatic limit) is employed to obtain the effective electronic Hamiltonian, which is used later on throughout the study to compute the spectral and topological properties. Particularly, in FOTPT, as we tune the e-p coupling, the system shows a topological-trivial insulator transition for $0<\alpha\leq 0.5$ and a trivial-topological-trivial transition for $0.54<\alpha<1.0$ associated with an SM-topological-trivial transition for $0.51<\alpha<0.54$. The formation of these topologically distinct phases is accompanied by bulk gap closing transitions at critical e-p coupling strengths, namely $\lambda_c$, $\lambda_{c_1}$, and $\lambda_{c_2}$. Nevertheless, there exists a gap in the bulk spectra below (topological) and above (trivial) $\lambda_c$ for $0<\alpha\leq 0.5$, while for $0.54<\alpha<1.0$, the same sustains in $\lambda<\lambda_{c_1}$ (trivial), $\lambda_{c_1}<\lambda<\lambda_{c_2}$ (topological), and $\lambda>\lambda_{c_2}$ (trivial) regimes, whereas it provides similar scenarios for $0.51<\alpha<0.54$ in the last two regimes of $\lambda$. Hence, apart from executing a TPT, the e-p interaction has been able to generate topologically nontrivial phases in the $\lambda_{c_1}<\lambda<\lambda_{c_2}$ regime for intermediate to higher range of $\alpha$ ($0.5<\alpha<1.0$), which are already present for lower range of $\alpha$ ($0<\alpha\leq 0.5$) in the absence of e-p coupling. These nontrivial phases induced by polarons are the QSH phases of an $\alpha$-$T_3$ lattice protected by TRS. The signature of the topological (trivial) phase is assured by the emergence (disappearance) of the QSH helical edge modes characterized by a non-zero (zero) $\mathbb{Z}_2$ invariant (or equivalently an odd/even number of crossings in HWCCs evolving through the half-BZ), as discussed elaborately in Secs.~\ref{textZ2}-\ref{textedge}. Finally, we summarize our findings for FOTPT in the form of a phase diagram (presented in Sec.~\ref{textphase}) which certifies that an $\alpha$-$T_3$ lattice hosts QSH phases and relative transitions caused solely by the e-p coupling.

Later, we introduce an in-plane magnetic field in the system to trigger the formation of a second order topological phase and to subsequently study the effect of e-p coupling on the emergent higher order states.
The magnetic field breaks TRS which is a key ingredient supporting the topology of the QSH insulator.
Simultaneously, the dispersive edge states of the zigzag nanoribbon gap out.
However, we observe that in-gap states with probability density confined at specific corners of an open boundary geometry emerge immediately as the magnetic field is introduced.
Importantly, these states confine themselves at those corners of a supercell where the two zigzag edges of the $\alpha$-$T_3$ lattice intersect.
This confinement is a consequence of the fact that the in-plane field is incapable of gapping out the original first order states on an armchair boundary, while being able to do so only on a zigzag one, inferring the inability of the magnetic field to disturb the inherent bulk topology of the system.
We verify this claim by employing the projected spin Chern number which shows a non-trivial value even in the presence of the magnetic field.
We further study the dependence of the real energy spectra on the e-p coupling $\lambda$.
It is observed that the in-gap states persist in the system till a critical value of the coupling parameter ($\lambda_c^{\mathrm{SOTI}}$) before completely merging into the bulk.
The numerical value of $\lambda_c^{\mathrm{SOTI}}$ is highly dependent on the strength of the in-plane field, whereas the dependence on the B-C coupling parameter $\alpha$ is weak.
Finally, we focus on the bulk bandstructure which shows a gap closing transition near $\lambda_c^{\mathrm{SOTI}}$, establishing a clear correspondence between the bulk and the boundary at the point where the second-order topology vanishes in the system. Thereafter, we mention the role of $\alpha$ and hence the interplay between the in-plane field and $\alpha$ on inducing the SOTI phases in an $\alpha$-$T_3$ lattice.    

\appendix
\section{Derivation of the ${k}$-space polaronic Hamiltonian}\label{text kspace Hamiltonian}
In the continuum limit, the Fourier transformed Hamiltonian can be obtained in a spinful tri-atomic sublattice basis, that is, in $({A_\uparrow~~B_\uparrow~~C_\uparrow~~A_\downarrow~~B_\downarrow~~C_\downarrow})$ basis, as
\begin{equation*}
\tilde{\mathcal{H}}_{\text{eff}}(\bf{k})=~~~~~~~~~~~~~~~~~~~~~~~~~~~~~~~~~~~~~~~~~~~~~~~~~~~~~~~~~~~~~
\end{equation*}
\begin{equation}
\begin{pmatrix}
\rho_{\sigma\sigma}^{11}(\bf k) & \rho_{\sigma\sigma}^{12}(\bf k) & 0 & 0 & \gamma_{\sigma\sigma^\prime}^{12}(\bf k) & 0\\\\
\rho_{\sigma\sigma}^{21}(\bf k) & \rho_{\sigma\sigma}^{22}(\bf k) & \rho_{\sigma\sigma}^{23}(\bf k) & \gamma_{\sigma\sigma^\prime}^{21}(\bf k) & 0 & \gamma_{\sigma\sigma^\prime}^{23}(\bf k)\\\\
0 & \rho_{\sigma\sigma}^{32}(\bf k) & \rho_{\sigma\sigma}^{33}(\bf k) & 0 & \gamma_{\sigma\sigma^\prime}^{32}(\bf k) & 0\\\\
0 & \gamma_{\sigma^\prime\sigma}^{12}(\bf k) & 0 & \rho_{\sigma^\prime\sigma^\prime}^{11}(\bf k) & \rho_{\sigma^\prime\sigma^\prime}^{12}(\bf k) & 0\\\\
\gamma_{\sigma^\prime\sigma}^{21}(\bf k) & 0 & \gamma_{\sigma^\prime\sigma}^{23}(\bf k) & \rho_{\sigma^\prime\sigma^\prime}^{21}(\bf k) & \rho_{\sigma^\prime\sigma^\prime}^{22}(\bf k) & \rho_{\sigma^\prime\sigma^\prime}^{23}(\bf k)\\\\ 
0 & \gamma_{\sigma^\prime\sigma}^{32}(\bf k) & 0 & 0 & \rho_{\sigma^\prime\sigma^\prime}^{32}(\bf k) & \rho_{\sigma^\prime\sigma^\prime}^{33}(\bf k)
\end{pmatrix},
\label{Ham:momentum space}    
\end{equation}
where $\rho$-blocks stand for the NN and NNN hopping (for bare KM model) elements, while $\gamma$-blocks refer to the RSOC terms which can be explicitly expressed as
\begin{subequations}
\begin{gather}
{\rho_{\sigma\sigma(\sigma^\prime\sigma^\prime)}^{11}(\bf k)}=\mathcal{M}-g^2\omega_0\pm\frac{2\tilde{t}_{so}Im(f_{\bf k})}{3\sqrt{3}}\cos\xi,\\
{\rho_{\sigma\sigma(\sigma^\prime\sigma^\prime)}^{12}(\bf k)}=-\tilde{t}(h_x^{\bf k}-ih_y^{\bf k})\cos\xi=\biggl[{\rho_{\sigma\sigma(\sigma^\prime\sigma^\prime)}^{21}(\bf k)}\biggr]^*,\\
{\rho_{\sigma\sigma(\sigma^\prime\sigma^\prime)}^{22}(\bf k)}=-g^2\omega_0\mp\frac{2\tilde{t}_{so}Im(f_{\bf k})}{3\sqrt{3}}(\cos\xi-\sin\xi),\\
{\rho_{\sigma\sigma(\sigma^\prime\sigma^\prime)}^{23}(\bf k)}=-\tilde{t}(h_x^{\bf k}-ih_y^{\bf k})\sin\xi=\biggl[{\rho_{\sigma\sigma(\sigma^\prime\sigma^\prime)}^{32}(\bf k)}\biggr]^*,\\  
{\rho_{\sigma\sigma(\sigma^\prime\sigma^\prime)}^{33}(\bf k)}=-\mathcal{M}-g^2\omega_0\mp\frac{2\tilde{t}_{so}Im(f_{\bf k})}{3\sqrt{3}}\sin\xi,
\end{gather} 
\label{matrho}
\end{subequations}
and
\begin{subequations}
\begin{gather}
{\gamma_{\sigma\sigma^\prime}^{12}(\bf k)}=-i\tilde{\beta}_R u_{\bf k}\cos\xi=\biggl[{\gamma_{\sigma^\prime\sigma}^{21}(\bf k)}\biggr]^*,\\
{\gamma_{\sigma\sigma^\prime}^{21}(\bf k)}=i\tilde{\beta}_R v_{\bf k}\cos\xi=\biggl[{\gamma_{\sigma^\prime\sigma}^{12}(\bf k)}\biggr]^*,\\   
{\gamma_{\sigma\sigma^\prime}^{23}(\bf k)}=i\tilde{\beta}_R u_{\bf k}\sin\xi=\biggl[{\gamma_{\sigma^\prime\sigma}^{32}(\bf k)}\biggr]^*,\\   
{\gamma_{\sigma\sigma^\prime}^{32}(\bf k)}=-i\tilde{\beta}_R v_{\bf k}\sin\xi=\biggl[{\gamma_{\sigma^\prime\sigma}^{23}(\bf k)}\biggr]^*,    
\end{gather}  
\label{matgamma}  
\end{subequations}
where $\sigma$ ($\sigma^\prime$) stands for the up (down) spin index, the angle $\xi$ is related to the parameter $\alpha$ as $\xi=\tan^{-1}\alpha$. The $ij$ components of the $\rho_{\sigma\sigma}$-block differ only by a relative sign in $\tilde{t}_{so}$ terms to those of the $\rho_{\sigma^\prime\sigma^\prime}$-block, while the $ij$ components of the $\gamma_{\sigma\sigma^\prime}$-block are complex conjugate to the $ji$ components of the $\gamma_{\sigma^\prime\sigma}$-block. The reduced hopping strengths (Holstein amplitudes), namely, $\tilde{t}$, $\tilde{t}_{so}$, and $\tilde{\beta}_R$ are expressed in Eq.~\ref{Holstein amp}. $h_x^k$, $h_y^k$ and $f_{k}$ in Eq. \eqref{matrho} and $u_k$ and $v_k$ in Eq. \eqref{matgamma} are given as
\begin{subequations}
\begin{equation}
h_x^{\bf k}=\sum_{i=1}^{3} {\cos(\bf{k.d_{i}})},~{h_y^{\bf k}= \sum_{i=1}^{3} \sin(\bf{k}.\bf{d_{i}})},~f_{\bf k}=\sum_{i=1}^{3} e^{(i\bf{k}.\bf{a_{i}})},
\label{hf}   
\end{equation}
\begin{equation}
u_{\bf k}(v_{\bf k})=-e^{-ik_ya_0}+\biggl[\cos\biggl(\frac{\sqrt{3}a_0k_x}{2}\biggr)\mp\sqrt{3}\sin\biggl(\frac{\sqrt{3}a_0k_y}{2}\biggr)\biggr]e^{\frac{ik_ya_0}{2}},   
\label{uv}
\end{equation}    
\end{subequations}
where the NN $\bf d$-vectors and the NNN $\bf a$-vectors have been mentioned in the caption of Fig.\,\ref{fig:model}. 
\label{appendix}

\section*{Acknowledgments}
K.B. sincerely thanks the Science and Engineering Research Board (SERB), Govt. of India, for providing financial support through the National Post Doctoral Fellowship (NPDF). S.L. acknowledges the Ministry of Education (MOE), Govt. of India, for providing financial support for her research work through the Prime Minister’s Research Fellowship (PMRF) May 2022 scheme. M.I. sincerely acknowledges the financial support from the Council of Scientific and Industrial Research (CSIR), Govt. of India to pursue this work.

\bibliography{references}
\end{document}